%% file: main.tex
\documentclass[11pt,a4paper]{article}
\input{preamble}
\begin{document}
\maketitle
\input{abstract}
\input{sections/introduction}
\input{sections/model}
\input{sections/results}

\input{sections/mechanisms}
\input{sections/related}
\input{sections/discussion}
\label{front-end}
\clearpage
\bibliographystyle{plain}
\bibliography{references}
\clearpage
\appendix
\input{proofs/navigation}
\input{proofs/capacity}
\input{proofs/finite_optimization}

\input{proofs/count_sampling}

\input{proofs/capacity_upper}

\input{proofs/hardness_gadget}
\input{proofs/bare_hardness}
\input{proofs/value_hardness}

\input{proofs/forest}
\input{proofs/single_chain}
\input{proofs/path}
\input{proofs/disclosure}
\end{document}

%% file: preamble.tex
\usepackage[T1]{fontenc}
\usepackage{lmodern}
\usepackage[margin=1in]{geometry}
\usepackage{amsmath,amssymb,amsthm,mathtools}
\usepackage{microtype,booktabs,array,longtable,enumitem}
\usepackage[colorlinks=true,linkcolor=blue,citecolor=blue,urlcolor=blue]{hyperref}
\usepackage{bbding}
\setlist{itemsep=2pt,topsep=4pt}
\newtheorem{theorem}{Theorem}[section]
\newtheorem{lemma}[theorem]{Lemma}
\newtheorem{proposition}[theorem]{Proposition}

\theoremstyle{definition}

\theoremstyle{remark}

\DeclareMathOperator{\conv}{conv}

\DeclareMathOperator{\supp}{supp}
\DeclareMathOperator{\Var}{Var}

\newcommand{\E}{\mathbb E}
\newcommand{\R}{\mathbb R}
\newcommand{\F}{\mathcal F}
\newcommand{\ind}{\mathbf 1}

\newcommand{\bits}{\ell}

\title{Counting Is Easier than Optimizing Exponential Bounds\\in Ordered Quota Systems}
\author{Ruiqing Yan \quad Yuxiang Zhao \quad Ruiting Wang \quad Ting Gan\,\textsuperscript{\Envelope} \quad Hao Huang\,\textsuperscript{\Envelope}\\[0.4em]
  \small School of Computer Science, Wuhan University, Wuhan, China\\[0.2em]
  \small 
  {\{yanruiqing,yuxiangzhao,wangrt,ganting,haohuang\}@whu.edu.cn}
  }
\date{}
\hypersetup{
    pdftitle={Counting Is Easier than Optimizing Exponential Bounds in Ordered Quota Systems},
    pdfauthor={Ruiqing Yan; Yuxiang Zhao; Ruiting Wang; Ting Gan; Hao Huang}
}

%% file: abstract.tex
\begin{abstract}
We compare the complexity of approximating a feasible mass and its optimal additive exponential upper bound on the same succinct input. Each ordered group contributes a fixed number of labels. The activities and all covering scores are nondecreasing along the supplied group orders. The bound is the family capacity of the feasible incidence hull, with its partition function taken over the complete quota source; a separation oracle for the hull is not supplied.

For every fixed supplied exchange rank, we give a deterministic fully polynomial-time approximation scheme (FPTAS) for the feasible mass and an exact weighted sampler with expected polynomial bit time on nonempty inputs. The number of covering rows can grow. At supplied and minimum representation rank two, the hardness exponent depends on the quotas. Write $s=1$ for binary chains and $s=2$ for unrestricted quotas. Finding an $N^\gamma$-approximate potential is NP-hard under deterministic polynomial-time Turing reductions for each fixed $0<\gamma<s$, where $N$ counts original labels. The reduction decodes the rational coordinates alone, without a support threshold or an objective value. A single query gives hardness for symmetric value approximation within $N^\gamma$ for each $0<\gamma<s/2$. The corresponding upper bounds have factors $O(N^s)$ and $O(N^{s/2})$.

Incidence forests admit capacity and mass FPTASs when each local quota domain is polynomially enumerable or its incident profiles share one affine scalar. A single half-quota chain of local rank two already gives an inverse-cubic precision barrier. A path of such blocks has $2^{N^\alpha}$-factor hardness for each fixed $0<\alpha<1$, with growing global rank. The hard paths still admit a mass FPTAS and exact sampling.
\end{abstract}

\begin{center}
\begin{minipage}{0.9\textwidth}
\end{minipage}
\end{center}

%% file: sections/introduction.tex
\section{Introduction}
An exponential upper bound on a constrained mass uses a partition function over the complete source and an inequality valid on every feasible word. The partition function can be easy to evaluate even when the feasible set has a complicated convex hull. We ask whether an algorithm that can approximate the constrained mass can also find a nearly optimal bound when every valid additive inequality is available in principle.

We study a product of fixed-quota choices. Each ordered group $C_c$ contributes exactly $g_c$ distinct labels. The positive activities and all covering scores are nondecreasing along the supplied order in each group. Let $\Omega$ be the complete quota source, $\F$ the family satisfying the covering rows, and $Z=\sum_{X\in\F}w(X)$ its mass. An additive field $\beta\in\R^N$ defines the partition function and tight feasible threshold
\[
 P(\beta)=\sum_{X\in\Omega}w(X)e^{\beta\cdot\chi(X)},\qquad
 m_\F(\beta)=\min_{X\in\F}\beta\cdot\chi(X).
\]
The optimal bound is
\begin{equation}\label{eq:intro-capacity}
 U=\inf_{\beta\in\R^N}P(\beta)e^{-m_\F(\beta)}.
\end{equation}
Fields can be signed and can distinguish labels with identical supplied scores. This is the family capacity of Straszak and Vishnoi~\cite{StraszakVishnoi2017} for the complete source polynomial and the hull $K=\conv\{\chi(X):X\in\F\}$. The supplied rows specify the discrete feasible words. Optimizing over all valid field--threshold pairs makes the bound depend on $K$.

For a full-source exponential proposal $Q_\beta(X)=w(X)e^{\beta\cdot\chi(X)}/P(\beta)$, the optimal rejection envelope for the target law $w(X)/Z$ on $\F$ is $P(\beta)e^{-m_\F(\beta)}/Z$. Thus $U/Z$ is the infimum of the expected rejection trials over this proposal class. In particular, on binary groups this class consists of all positive independent Bernoulli proposals. The capacity is a benchmark for certificates and proposals of this form. Its value can differ from $Z$, and other exact sampling algorithms can be much faster.

\paragraph{Example.}
Three independent on/off generators have ratings $2,2,1$ and must meet demand $3$. Use binary groups with unit activities, and let $x_i$ indicate that generator $i$ is on. Feasibility is equivalent to having at least two generators on, so $Z=4$ and the feasible probability under the fair source is $1/2$. Suppose an exact conditional sampler uses a proposal that can change the three on-probabilities independently. Its optimal expected number of proposals is $U/Z$.

The valid inequality $\sum_i x_i\ge2$, with a field of $\log 2$ in each high coordinate, gives a bound of $3^3/2^2=27/4$. This is optimal over all fields. Independent Bernoulli variables of mean $2/3$ have the same original-label marginals as the uniform law on the three states with exactly two generators on. Their entropy is $\log(27/4)$, which gives the matching lower bound by Gibbs' inequality. The optimal proposal uses on-probability $2/3$ for each generator and takes $27/16$ trials in expectation.

Let $a=(2,2,1)$, and let $h$ denote binary entropy in nats. Every field obtained by multiplying the supplied demand row by $u\ge0$ has tight threshold $3u$. Independent Bernoulli variables of mean $3/5$ have expected score $3$. Gibbs' inequality then gives
\[
 \log P(ua)-3u\ \ge\ 3h(3/5)\ >\ 3h(2/3)\ =\ \log(27/4),
\]
where $ua$ is written in the high coordinates. Thus even the best tilt along the physical demand score performs strictly worse within this proposal class.

\paragraph{Results.}
For every fixed supplied exchange rank, we give a deterministic FPTAS for $Z$ and an exact weighted sampler on nonempty inputs. Exchange rank measures the dimension of within-group score differences. The count algorithm compresses the simultaneous suffix tests in this dimension and allows arbitrarily many supplied rows.

Already at supplied and minimum representation rank two, the polynomial approximation exponents for family capacity are tight on the same input class. We distinguish a bare field, a field with a valid threshold and an upper value, and a symmetric approximation to $U$. A gap in $U$ alone would leave open whether a good field could be found without evaluating its tight threshold. Our search reduction resolves this issue by encoding several possible feasible boundary families whose explicit original-label means average to one high-entropy source mean. Comparing rational linear forms in the returned field identifies a candidate that can safely be discarded. Repeated pruning decides the original instance using neither a threshold nor an objective value. This coordinate decoder is the additional ingredient needed to separate counting from the search for an optimal exponential bound.

Incidence forests give a second comparison. When the local quota domains are enumerable or their incident profiles share one affine scalar, the forest supports both approximation tasks. A single long chain with two independent profiles yields a capacity obstruction at inverse-cubic accuracy. A path of such blocks amplifies it to a stretched-exponential factor, while the same path retains a mass FPTAS and an exact sampler. The path's global rank grows, so the count and sampler on these instances use a separate two-state decomposition.

%% file: sections/model.tex
\section{Model and output contracts}\label{sec:model}
The disjoint groups $C_1,\ldots,C_d$ are ordered chains of explicitly listed labels. Write $n_c=|C_c|$, $0\le g_c\le n_c$, and
\[
 \Omega=\prod_{c=1}^d\binom{C_c}{g_c},\qquad N=\sum_c n_c.
\]
A word $X$ has incidence vector $\chi(X)\in\{0,1\}^N$ and weight $w(X)=\prod_{i\in X}\lambda_i$. Activities $\lambda_i$ are positive rationals. For $j\in[p]$, the rational row $a_j$ and threshold $\theta_j$ define
\[
 \F=\{X\in\Omega:a_j\cdot\chi(X)\ge\theta_j\text{ for all }j\},
 \qquad Z=\sum_{X\in\F}w(X).
\]
The case $N=0$ has only the empty word, of weight one, and is
handled directly by the constant row tests. All numerical bounds
below use $N\ge1$.
Each $a_j(i)$ and $\lambda_i$ is nondecreasing along each chain. Scores and thresholds may be signed. All input rationals are encoded in binary. The total input length $\bits$ includes the explicit labels and quotas. The number $p$ is part of the input. The simultaneous top quota suffix decides emptiness. We state capacity results for nonempty $\F$ and recognize emptiness separately. Logarithms are natural.

Every additive valid inequality $b\cdot\chi(X)\ge t$ gives the upper bound $P(b)e^{-t}\ge Z$. A finite collection with nonnegative multipliers combines into one such pair. Consequently the infimum over all finite valid collections equals~\eqref{eq:intro-capacity}. No new inequalities have to preserve the supplied common order. For a family $A\subseteq\Omega$, write $U(A)$ and $m_A$ for the same definitions with $\F$ replaced by $A$, and put
\[
 G_A(\beta)=\log P(\beta)-m_A(\beta),\qquad G=G_\F.
\]
Infima need not be attained by finite potentials.

\paragraph{Supplied and minimum rank.}
Delete frozen chains only when forming exchange coordinates. The matrix whose columns are $(a_j(v)-a_j(u))_{j\in[p]}$ for pairs in a nonfrozen chain has rank $q(I)$. Equivalently, it is the rank of the score rows modulo chainwise constants. Gaussian elimination computes this supplied rank. The minimum representation rank $q_{\min}(\F)$ minimizes this rank over finite exact common-order covering representations of $\F$ on the same original quota source, permitting another common order. Algorithms here receive a representation; they do not compute $q_{\min}$. Our rank-two lower bounds supply a rank-two representation and additionally guarantee $q_{\min}=2$.

The local exchange rank of a chain is the rank of its incident score profiles modulo constants on that chain. The row--chain incidence graph joins a row to a nonfrozen chain exactly when that restriction is nonconstant. These are properties of the supplied representation.

\begin{samepage}
\paragraph{Approximation tasks.}
For a factor $\rho\ge1$ we distinguish the following polynomial-bit rational outputs:
\begin{center}
\begin{tabular}{@{}>{\raggedright\arraybackslash}p{.20\linewidth}p{.73\linewidth}@{}}
\toprule
Output & Required guarantee\\
\midrule
Bare potential & $\beta$ with $P(\beta)e^{-m_\F(\beta)}\le\rho U$. No threshold is returned.\\[3pt]
Certified upper bound & $(\beta,t,V)$ with $t\le m_\F(\beta)$ and $U\le P(\beta)e^{-t}\le V\le\rho U$.\\[3pt]
Symmetric value & $V$ with $U/\rho\le V\le\rho U$.\\
\bottomrule
\end{tabular}
\end{center}
\end{samepage}
A deterministic mass FPTAS returns $\widehat Z$ with $Z/(1+\varepsilon)\le\widehat Z\le Z$, in time polynomial in $\bits$ and $1/\varepsilon$. We take $0<\varepsilon\le1$; a larger requested tolerance may be replaced by one. Exact sampling means the law $w(X)/Z$ on the original labels, using fair random bits with expected polynomial running time. Throughout, ``fixed rank'' permits the polynomial degree to depend on rank; it is not a claim of fixed-parameter running time $f(q)\bits^{O(1)}$.

%% file: sections/results.tex
\section{Main results}\label{sec:results}
We first give a counting and sampling result for the input class used in the capacity hardness reductions.

\begin{theorem}[Counting at fixed supplied rank]\label{thm:count-main}
For every fixed $q$, ordered quota systems with $q(I)\le q$ admit a deterministic mass FPTAS. Emptiness is detected separately. On nonempty instances, exact weighted sampling takes expected polynomial bit time. The number of covering rows may grow with the input. No bound on binary-encoded score magnitudes or activity ratios is required.
\end{theorem}

The activity order is used in the simultaneous repair argument. It is not needed by all positive capacity algorithms below. The proof of Theorem~\ref{thm:count-main} gives a rounded superset with relative mass at most $(1+N^2)^q$ and a deterministic compression whose dimension is controlled by $q$, rather than by $p$.

\begin{theorem}[The rank-two separation]\label{thm:separation-main}
Let $s=1$ for the class in which every nonfrozen chain selects or omits one label, and let $s=2$ for unrestricted quotas.
\begin{enumerate}[label=(\roman*)]
\item On supplied rank two, polynomial bit time suffices for a certified upper approximation of factor $128N^s$ and a symmetric value approximation of factor $O(N^{s/2})$. Supplied rank at most one admits a certified FPTAS.
\item For every fixed $0<\gamma<s$, a polynomial-time algorithm returning an $N^\gamma$-approximate bare potential would imply $\mathrm P=\mathrm{NP}$. The reduction is a deterministic polynomial-time Turing reduction. It uses unit activities, two strictly increasing integer score rows, and instances with $q_{\min}(\F)=2$ on both decision outcomes. For $s=1$ the chains have two labels; for $s=2$ all nonfrozen chains are half-quota chains.
\item For every fixed $0<\gamma<s/2$, the same intrinsic rank-two classes have NP-hard symmetric value approximation within $N^\gamma$, already by a single oracle query. A single-query reduction also gives hardness of certified upper bounds within $N^\gamma$ for every fixed $0<\gamma<s$.
\end{enumerate}
Theorem~\ref{thm:count-main} applies to all these hard inputs.
\end{theorem}

Thus the powers in the upper and lower bounds agree. We do not assert optimal constants or exclude subpolynomial improvements. Dividing a certified upper estimate by the square root of its factor gives a symmetric estimate with exponent $s/2$. A bare potential has no efficiently available tight threshold, and its lower bound needs an independent decoder.

\paragraph{Incidence forests.}
A chain is \emph{scalar} if there are rational $s_i$ and intercepts $b_{jc}$ and nonnegative slopes $\alpha_{jc}$ such that
\begin{equation}\label{eq:scalar}
 a_j(i)=b_{jc}+\alpha_{jc}s_i\qquad(i\in C_c)
\end{equation}
for every incident row. This is an affine condition, stronger than merely sharing an order. It can be checked by choosing any nonconstant incident profile and testing proportional differences. Alternatively, a chain may be handled by listing every quota word, keeping its original labels and scores. Let $D$ be the number of enumerated local words summed over those chains. A list can be generated by enumerating quota subsets and computing exact scores in time polynomial in $\bits,D$.

\begin{theorem}[Computing on forests]\label{thm:forest-main}
Suppose the row--chain incidence graph is a forest and each chain is scalar or its complete local domain is enumerated. There are deterministic algorithms for certified capacity and mass within factor $1+\varepsilon$, polynomial in $\bits,D,1/\varepsilon$, and, on nonempty instances, an exact weighted sampler in expected time polynomial in $\bits,D$. There is no rank or degree bound. If $D$ has a uniform polynomial bound in the original input, both approximation algorithms are FPTASs on that input, and exact sampling takes expected polynomial bit time.

For binary chains, this capacity boundary is sharp in incidence treewidth: width one has a certified FPTAS, whereas width exactly two has bare-potential hardness under deterministic polynomial-time Turing reductions within $N^\gamma$ for every fixed $0<\gamma<1$.
\end{theorem}

In particular, selecting or omitting one label always gives an enumerable domain, without any affine-profile assumption. For long intermediate-quota chains the local condition matters, even on a path.

\begin{theorem}[One long chain and a path]\label{thm:path-main}
There is an absolute $c>0$ with the following properties.
\begin{enumerate}[label=(\roman*)]
\item On one unit-activity half-quota chain of $n$ labels, two strictly increasing integer rows make factor $1+c/n^3$ approximation NP-hard for each of the three output contracts. One oracle query suffices, and both outcomes have minimum representation rank two.
\item For every fixed $0<\alpha<1$, factor $2^{N^\alpha}$ approximation is NP-hard for each contract on unit-activity systems whose row--chain incidence graph is a single path. Each private half-quota chain has local rank two; intervening quota-one chains have local rank one. With $t$ private blocks, the minimum global representation rank is $2t$ on both outcomes.
\end{enumerate}
The path lower bound also uses a single oracle query. On the instances produced by both reductions, the true mass admits a deterministic FPTAS and the uniform law on feasible labelled words admits an exact sampler in expected polynomial bit time. All size and approximation factors count original labels, including the path's gate labels.
\end{theorem}

The second assertion does not follow by treating the growing global rank as fixed. Its positive algorithms use a separate two-state path decomposition and recursive rejection. The single-chain statement concerns inverse-polynomial accuracy; constant-factor capacity approximation on one arbitrary long chain is left open.

\begin{table}[t]
\centering\small
\begin{tabular}{@{}>{\raggedright\arraybackslash}p{.29\linewidth}>{\raggedright\arraybackslash}p{.40\linewidth}>{\raggedright\arraybackslash}p{.23\linewidth}@{}}
\toprule
Capacity input class & Hard output and factor & Decision reduction\\
\midrule
Minimum rank two & Bare potential at $N^\gamma$, $0<\gamma<s$ & Polynomial-time Turing\\
Minimum rank two & Certified upper at $N^\gamma$, $0<\gamma<s$ & One query\\
Minimum rank two & Symmetric value at $N^\gamma$, $0<\gamma<s/2$ & One query\\
One half-quota chain, minimum rank two & All three at $1+c/n^3$ & One query\\
Path with local ranks at most two and minimum global rank $2t$ & All three at $2^{N^\alpha}$, $0<\alpha<1$ & One query\\
\bottomrule
\end{tabular}
\caption{Capacity approximation lower bounds. The exponents $\gamma$ and $\alpha$ are fixed. Here $s=1$ for binary hard instances and $s=2$ for the unrestricted-quota hard instances. All these families have deterministic mass FPTASs and exact samplers.}
\label{tab:capacity-hardness}
\end{table}

\subsection{Bit complexity}\label{sec:complexity}
The bounds below count bit operations and bits of storage. Write $\bits$ for the binary input length, including the encoding of a rational tolerance when one is supplied. The parameters are the number $p$ of covering rows, the supplied exchange rank $q$, and, for forests, the total enumerated domain size $D$. Each table entry and exact arithmetic operation below has bit cost polynomial in $\bits$ and in the bit lengths of the displayed indices. In particular, the bounds do not involve the numerical magnitudes of the scores.

\paragraph{Fixed-rank mass and sampling.}
Let $r\le q$ be the rank of the normalized positive-budget charge matrix. The case $r=0$ uses only quotas and pins. For $r>0$ put
\[
 K_0=(1+N^2)^r,\qquad H=8rN^2,\qquad
 M=\left\lceil 4(p+1)(N+1)^2 K_0\,\frac{2+\varepsilon}{\varepsilon}\right\rceil.
\]
There are $S=O((N+1)^2(H+1)^r)$ generator states over all layers. Enumerating the row bases takes at most $(1+p)^r$ determinant tests. Exact suffix masses on this generator give the sampler, with at most $K_0$ expected proposals. Both expected time and space are bounded by $\operatorname{poly}(\bits)((1+p)^r+S K_0)$, with space bounded also by $\operatorname{poly}(\bits) S$ when bases are tested one at a time. The deterministic count stores $O(Sp(M+1))$ suffix CDF entries and visits at most $S(1+pM)^{O(r)}$ representative prefixes. Both its time and space are bounded above by
\[
 \operatorname{poly}(\bits,\log M)
 \bigl[(1+p)^r + S\bigl(p(M+1)+(1+pM)^{O(r)}\bigr)\bigr].
\]
In particular the polynomial degree can depend on $q$. The merges change the acceptance probability by at most $2pN^2/M$. The lower bound $1/K_0$ on that probability gives relative error at most $2pN^2 K_0/M$ before the final one-sided rescaling.

\paragraph{Capacity.}
At internal tolerance $0<\tau\le1$, the finite optimizer uses a box of radius $R=O(\bits+\log(1/\tau))$ and $O(1+N^2 R^2/\tau^2)$ calls to a support routine at additive tolerance $\delta=\tau/16$. At each iteration it evaluates the full-source marginals using elementary symmetric polynomials. It stores the current field, an accumulating average, and reusable workspaces for source evaluation and support. Define the cost index bound $C=1+\lceil N^2R/\delta\rceil$. At rank one an oracle uses a quota-and-cost table with $O((N+1)^2 C)$ entries. The fixed-rank approximation uses $N^{O(q)}$ charts and $N^{O(q)}C$ entries and row tests in total per call, up to a factor polynomial in $\bits$. It takes constant internal relative tolerance for the stated rank-two factors. Forest support uses scalar quota-and-cost tables together with row convolutions on the cost range and enumerated states. For a coarse bound, each forest call uses at most $\operatorname{poly}(\bits)(1+D+C)^2$ bit time and space. The certified forest and rank-one FPTASs take $\tau=\varepsilon/4$. The runtime is the call count times the per-call and source-evaluation costs, and the workspace can be reused. See Appendices~\ref{comp:finite}, \ref{comp:upper}, and~\ref{app:forest}.

\paragraph{Forest mass and sampling.}
Write $E$ for the number of row--chain edges. With trimming tolerance $\eta=\varepsilon/[2(1+N+E)]$, a trimmed tail has $B_0=O(1+\bits/\eta)$ entries. Scalar local profiles have at most $B_0$ states each. The remaining local profiles have $D$ states in total. A row convolution enumerates pairs of entries from its two lists before trimming, and conditional row values are computed at the parent states by queries to one trimmed tail. Thus a coarse bound for both time and space is $\operatorname{poly}(\bits)(1+D+B_0)^2$. The sampler uses one initial mass call, then at most $2N$ residual mass calls per trial at tolerance $1/(4N)$, with at most four expected trials. Residual calls use the same bounds with the residual domains and the requested tolerance.

%% file: sections/mechanisms.tex
\section{Why the separation occurs}\label{sec:mechanisms}
The fixed-rank algorithms use enlarged word families. The hardness proofs use feasible original-label means.
Both keep the original labelled source (Appendix~\ref{app:guide}).

\subsection{Feasible support versus feasible mean}
Put $K=\conv\{\chi(X):X\in\F\}$. The family-capacity identity gives
\begin{equation}\label{eq:entropy-front}
 \log U=\max_{\mu:\,\E_\mu\chi\in K}
 \left(H(\mu)+\E_\mu\log w\right),
 \qquad
 \log Z=\max_{\mu:\,\supp\mu\subseteq\F}
 \left(H(\mu)+\E_\mu\log w\right).
\end{equation}
In the first maximum, the law may use any word in the complete source. Matching its inclusion
probabilities with those of a feasible law at \emph{every original
label} certifies a capacity lower bound. Matching only supplied row
scores would not control an arbitrary signed potential.
Appendix~\ref{app:capacity} includes boundary means and unattained infima.
Although $P$ is easy to evaluate, the feasible incidence hull need not
admit efficient support minimization.

\subsection{One rounded source for sampling and counting}
Use nonnegative top-word losses, with zero budgets enforced by pins.
A maximum-determinant \emph{row} basis rounds the normalized
positive-budget losses to a polynomial-state family $G\supseteq\F$.
A separate maximum-determinant \emph{column} basis repairs its words
with at most $r\le q$ disjoint upgrades. Activities do not decrease;
inverse-swap lists then give
\[
 Z\le Z_G\le K_0Z,\qquad K_0=(1+N^2)^r.
\]
Exact suffix masses sample $G$; testing the original rows gives the
exact feasible law within $K_0$ expected proposals, without computing
the repair map.

Deterministic counting compresses this acceptance calculation.
At a fixed generator state all prefixes share one suffix law.
Approximate one-row suffix CDFs define keys at the true remaining
budgets. For each fixed row, the events for two prefixes at the same key
are nested under that suffix law. A union bound over rows thus controls
the change in joint acceptance. Each occupied key extends one actual
representative, multiplying the key's entire mass by each transition
probability. Its successor keys use the extended representative's
exact remaining budgets. The forward invariant sums merge errors, and
$Z/Z_G\ge1/K_0$ converts them to relative counting error.
All prefix-charge vectors lie in a space of dimension at most $r$, so the CDF
breakpoint hyperplanes yield polynomially many keys at fixed $r$,
including equality faces and arbitrarily many rows
(Appendix~\ref{comp:count}).

\subsection{Certifying a capacity bound}
The positive capacity algorithms supply one interface:
\[
 X\in A,\qquad
 t\le m_A(\beta)\le\beta\cdot\chi(X)\le t+\delta.
\]
Finite numerical clipping and outward evaluation convert this
interface into rational outputs (Appendix~\ref{comp:finite}).
At fixed supplied rank bound $q\ge1$, charts keep one exchange coordinate exact and round
the others, defining a query-independent family $\F\subseteq D$.
At most $q-1$ upgrades repair each word. A distinct clipping argument
for \emph{effective activities} bounds $U(D)/U(\F)$ by
$O(N^{s(q-1)})$. The witness may lie in $D\setminus\F$, but its
threshold is valid because $t\le m_D\le m_\F$
(Appendix~\ref{comp:upper}). On forests, scalar or enumerated domains
yield support, trimmed mass, and exact rejection correction directly
for $\F$ (Appendix~\ref{app:forest}).

\subsection{Reading a decision from a bare potential}
A group with $d$ defects has $\binom cd^s$ labelled choices,
where $s=1$ for binary chains and $s=2$ for half-quota chains.
Two increasing rows encode a bulk and a boundary. Carry-free radix
digits force each boundary word to be a partition witness of one
color. An original-label incidence trade forces minimum rank two
on either outcome (Appendix~\ref{comp:gadget-section}).

For $K\ge2$ colors encoding $a$-item lists, define $\mu_r$ by mean defect $1/4$ in each
group of color $r$ and zero in the others. A solution and its
complement, balanced signs, and uniform labelled choices realize
$\mu_r$ feasibly whenever $r$ is soluble. The product Gibbs law $Q_z$ has entropy $\mathcal S$ and mean
$K^{-1}\sum_{r=0}^{K-1}\mu_r$. The average includes insoluble colors as well. Each group has mean defect
$1/(4K)$, distributed uniformly over its labels. If a soluble
color minimizes $\beta\cdot\mu_r$, then
\[
 m_\F(\beta)\le\beta\cdot\mu_r
 \le\beta\cdot\E_{Q_z}\chi,
 \qquad G(\beta)\ge\mathcal S.
\]
Deletion can change the answer only when it removes the \emph{unique} soluble color.
In that case a color-specific comparison field gives
\[
 \log U\le\mathcal S-\Delta,\qquad
 \Delta=(1-1/K)L-O_K(L^2/a),\quad L=s\log c+O_K(1).
\]
Lemma~\ref{lem:deletion-front} proves this gap. For every fixed
$0<\gamma<s$, polynomially growing $c$ and fixed $K$ can make
$\Delta>\gamma\log N$. Thus a promised $N^\gamma$-approximate
field cannot rank the unique soluble color lowest, even in a tie.

For the all-high word $T$, the actual decoder uses only
\[
 \beta\cdot\mu_r=\beta\cdot\chi(T)
 +\frac{\sum_{i\ {\rm low\ in\ color}\ r}\beta_i
       -\sum_{i\ {\rm high\ in\ color}\ r}\beta_i}{4c}.
\]
It compares rational color sums, without evaluating $G$, $m_\F$,
$z$, or $\Delta$. Branching and repeated deletion keep at most
$K-1$ lists while preserving their disjunction. Appendix~\ref{comp:bare-section}
proves the parameter choice and polynomial query and bit bounds;
Appendix~\ref{comp:value-section} gives the separate one-query value decoder.

\subsection{A surviving shell and its path amplification}
On one half-quota chain, let $G_0(X)=\sum_{i\in X}i$ be the position
grade. On YES, a complementary feasible pair certifies full capacity
$B=\binom n{n/2}$. On NO, a stencil deletes some first-shell words.
Adjacent-swap laws inside a guaranteed feasible sub-shell nevertheless
realize the optimal grade tilt's original-label means. Complement-reflection preserves grade and stencil sum, so it keeps
these shell witnesses feasible regardless of their partition statistic. The resulting exact identity is
\begin{equation}\label{eq:shell-front}
 \log(B/U)=v-K(v)=\frac{24+o(1)}{n^3},
 \qquad K(v)=\log\E e^{v(G_0-\E G_0)},\quad K'(v)=1.
\end{equation}
A finite gap and a second feasible mean give a rational centered-norm
decoder (Appendix~\ref{app:single-chain}).

Join $t$ copies by gates with one low and $m$ high labels. High
restores the private inequalities; low forces both neighbors top.
The all-high face gives
\[
 m^{t-1}U^t\le U_{\rm path}\le(m+1)^{t-1}U^t,
\]
so near-optimal joint fields have a small sum of nonnegative private
regrets. Choosing $m=n^4$ and polynomially many copies as
in~\eqref{eq:path-exponent} makes a strict majority decodable even
at factor $2^{N^\alpha}$ for any fixed $0<\alpha<1$. The paths use a separate two-state count
and balanced exact sampler, despite growing global rank
(Appendix~\ref{app:path}).

%% file: sections/related.tex
\section{Related work}
\paragraph{Family capacity and oracle access.}
Straszak and Vishnoi~\cite{StraszakVishnoi2017} define family capacity, identify it with a source entropy optimum subject to a mean constraint, and give an approximation algorithm (Lemma 3.6 in arXiv:1611.04548v1) with an evaluation oracle for the source polynomial and a separation oracle for the target incidence hull. The complete quota polynomial supplies the evaluation interface by elementary symmetric recurrences. The capacity algorithm allows arbitrary target families of the given cardinality. Their coefficient-sum approximation guarantees use additional matroid hypotheses. Our input instead specifies the target by covering rows on discrete quota words. Those rows need not describe their incidence hull. In the generator example the mean $(3/5,3/5,3/5)$ satisfies the supplied row and violates the feasible inequality $\sum_i x_i\ge2$. Our lower bounds apply to the succinct covering-row representation and include the search task of returning only a field.

\paragraph{Counting and maximum entropy.}
Singh and Vishnoi~\cite{SinghVishnoi2014} connect counting with maximum entropy on a fixed discrete support at a prescribed marginal. The theorem numbers here refer to their full version, arXiv:1304.8108v1. Their forward reductions (Theorems 2.6 and 2.8) require affine-hull equations, a relative-interiority radius, and total and single-coordinate restricted counts under external fields. Their reverse reduction (Definition 2.10 and Theorem 2.11) also uses separation over the marginal polytope. In their Appendix B, Theorem B.2 assumes separation and the KL-optimization oracle of Definition B.1, which compares objective values with a queried threshold; Theorem B.5 obtains near-optimal dual fields from generalized counting.

At a prescribed mean $x$, the weighted primal objective is $H(\nu)+\E_\nu\log w=H(\nu)+\sum_i x_i\log\lambda_i$. Thus exponential weighting adds a constant and leaves the fixed-mean primal optimizer unchanged (their Lemma B.6). Here $\log Z$ optimizes over laws supported on $\F$, whereas $\log U$ optimizes over complete-source laws whose means lie in $K$. Access to general reweightings and to the implicit hull remains essential to the cited reductions.

The mass FPTAS in this paper takes activities that share the supplied score order. A general external field changes them to $\lambda_i e^{\beta_i}$ and can reverse that order, even with rational effective activities. For a quota-one group with scores $0<1$ and unit base activities, effective activities $(2,1)$ already violate the hypothesis. The count guarantee therefore supplies neither the general reweighted counting interface nor a separation oracle for $K$. Straszak and Vishnoi~\cite[Theorem~7]{StraszakVishnoi2019} bound near-optimal dual fields in terms of the support hull's unary facet complexity and the weight bit length, uniformly over interior and boundary means. We use a quota-specific clipping argument and an additive support oracle to obtain a rational field and a valid rational threshold.

\paragraph{Counting and sampling methods.}
Gopalan et al.~\cite[Theorem~1.2]{GopalanEtAl2011} give a deterministic FPTAS for multidimensional knapsack with a fixed number of constraints using branching programs and small-space sources. For a single Boolean knapsack constraint, Feng and Jin~\cite{FengJin2025} improve the randomized running time, and Jin, Park, and Saha~\cite{JinParkSaha2026} improve the deterministic running time. Here the number of supplied rows can grow even at fixed exchange rank. Linearly dependent rows can all constrain the feasible words. The algorithm retains their tests and controls the number of joint keys by the dimension of the prefix charge space. Its bounded-coordinate basis argument is classical~\cite{AwerbuchKleinberg2004}, and prefix-cover compression has a close predecessor in Gopalan, Klivans, and Meka~\cite{GopalanKlivansMeka2012}.

Dyer~\cite{Dyer2003} uses rounded supersets and exact backtracking with rejection. We bound the acceptance probability by a quota-preserving repair using at most the exchange rank many upgrades. Rounded-cost optimization with category quotas also has a min-knapsack precedent~\cite{PatelKhanLouis2021}. These methods provide the counting and support-oracle ingredients. The complexity separation compares those algorithms with reductions for three different outputs of the same family-capacity problem.

%% file: sections/discussion.tex
\section{Scope and questions}
The rank-two polynomial exponents are settled under the stated contracts. At larger fixed rank, the constructive bound has exponent $s(q-1)$; determining its optimal exponent requires more than an independent product of rank-two gadgets. On one arbitrary long chain, the present inverse-cubic precision barrier leaves fixed-factor approximation open. On forests, the positive theorem identifies a sufficient local representation condition, and the path construction gives stretched-exponential hardness for a family of paths outside that class. A broader classification of tractable local domains remains open.

%% file: proofs/navigation.tex
\section{Guide to the complete proofs}\label{app:guide}
The appendices prove the statements on the original quota source.
The table lists direct dependencies and the interface each proof must
establish; none uses an unproved companion result.

\begin{center}
\small
\begin{tabular}{@{}>{\raggedright\arraybackslash}p{.24\linewidth}>{\raggedright\arraybackslash}p{.18\linewidth}>{\raggedright\arraybackslash}p{.50\linewidth}@{}}
\toprule
Proof & Uses & Main obligation\\
\midrule
Capacity identity (\ref{app:capacity}) & Finite Gibbs principle; minimax & Match every original-label marginal; allow boundary means and unattained infima.\\[3pt]
Finite optimization (\ref{comp:finite}) & Complete-source evaluation & Convert the additive support contract~\eqref{comp:support-contract} into rational $(\beta,t,V)$.\\[3pt]
Fixed-rank count (\ref{comp:count}) & Quota losses; basis bounds & Repair bounds acceptance; the common suffix law permits probability-key merges; \eqref{comp:forward-invariant} pays their total error.\\[3pt]
Rank upper bound (\ref{comp:upper}) & \ref{comp:finite}; quota losses of~\ref{comp:count} & Fix $D\supseteq\F$ independently of the query; prove repair, chart dominance, and $t\le m_D\le m_\F$.\\[3pt]
Shared gadget (\ref{comp:gadget-section}) & Exact integer arithmetic & Characterize bulk and corner; preserve all labels and force minimum rank two.\\[3pt]
Bare / value hardness (\ref{comp:bare-section}, \ref{comp:value-section}) & \ref{comp:gadget-section}; Gibbs; \ref{comp:count} for counting & Decode bare coordinates using labelled means; use an exact rational-power comparison for values. Only the bare decoder prunes a list.\\[3pt]
Forest algorithms (\ref{app:forest}) & \ref{comp:finite}; tree messages & Keep exact score tests for support; compare tails for mass; correct the original-label proposal exactly.\\[3pt]
Single chain (\ref{app:single-chain}) & \ref{app:capacity}; \ref{comp:count} for counting & Realize tilted label marginals on surviving shell words, then decode the finite gap by a rational norm test.\\[3pt]
Path (\ref{app:path}) & \ref{app:single-chain} & Use the all-high face for regret, product trades for rank, and separate transfer / renewal arguments for count / sampling.\\
\bottomrule
\end{tabular}
\end{center}

\paragraph{Reading routes.}
For the central separation, read the output contracts in
Section~\ref{sec:model}, then Appendices~\ref{comp:count},
\ref{comp:gadget-section}, and~\ref{comp:bare-section};
the last appendix proves the color deletion rule. For positive capacity algorithms, read
Appendix~\ref{comp:finite} before~\ref{comp:upper} or~\ref{app:forest}.
For the local obstruction and amplification, read~\ref{app:single-chain}
before~\ref{app:path}.

%% file: proofs/capacity.tex
\section{The original-source capacity identity}\label{app:capacity}
We state the classical family-capacity and entropy identities for our source and cut conventions~\cite{StraszakVishnoi2017}. They hold for any finite source with positive weights. No ordering assumption is needed.

Let $\Delta(\Omega)$ be the simplex of probability laws on $\Omega$, let $K=\conv\chi(\F)$, and set $\mathcal E(\mu)=H(\mu)+\E_\mu\log w$. For a finite valid block $b_h\cdot\chi(X)\ge t_h$ and multipliers $u_h\ge0$, put $\beta=\sum_hu_hb_h$ and $t=\sum_hu_ht_h$. Then $t\le m_\F(\beta)$, so its exponential sum is at least $P(\beta)e^{-m_\F(\beta)}$. Conversely the single signed inequality $\beta\cdot\chi(X)\ge m_\F(\beta)$ is valid for every finite $\beta$ and may be used with multiplier one. Taking infima proves~\eqref{eq:intro-capacity}.

\begin{proposition}[Entropy and labelled means]\label{prop:entropy}
For nonempty $\F$,
\[
 \log U=\max_{\substack{\mu\in\Delta(\Omega)\\\E_\mu\chi\in K}}\mathcal E(\mu).
\]
The maximum exists. In particular, if a source law $\mu$ and a feasible law $\nu$ satisfy $\E_\mu\chi_i=\E_\nu\chi_i$ for every label, then $\log U\ge\mathcal E(\mu)$.
\end{proposition}
\begin{proof}
For fixed $\beta$, the finite Gibbs principle states
\[
 \log P(\beta)=\max_{\mu\in\Delta(\Omega)}\bigl[\mathcal E(\mu)+\beta\cdot\E_\mu\chi\bigr].
\]
The difference is the nonnegative relative entropy from $\mu$ to $w(X)e^{\beta\chi(X)}/P(\beta)$. Hence
\[
 \log U=\inf_\beta\max_{(\mu,x)\in\Delta(\Omega)\times K}
 \bigl[\mathcal E(\mu)+\beta\cdot(\E_\mu\chi-x)\bigr].
\]
The maximizing domain is compact and convex; the integrand is continuous and concave there, and affine in $\beta$. Sion's one-compact-side minimax theorem~\cite{Sion1958} interchanges the two extrema. The inner infimum is $\mathcal E(\mu)$ when $\E_\mu\chi=x$ and is $-\infty$ otherwise, as seen by taking $\beta=-r(\E_\mu\chi-x)$ and sending $r$ to infinity. The feasible set of laws is nonempty and compact, and $\mathcal E$ is continuous. This proves existence and the formula. A law $\nu$ supported on $\F$ has its mean in $K$, giving the last assertion.
\end{proof}

The corresponding Gibbs principle on $\F$ gives $\log Z=\max_{\supp\mu\subseteq\F}\mathcal E(\mu)$. For fixed $x\in K$, the same minimax argument gives
\[
 \inf_{\beta}\bigl(\log P(\beta)-\beta\cdot x\bigr)
 =\max_{\mu:\E_\mu\chi=x}\mathcal E(\mu).
\]
Thus $U=\max_{x\in K}\inf_{z>0}\sum_Xw(X)z^{\chi(X)}/z^x$, precisely the family capacity. These formulas include boundary means and do not require finite attainment of the dual infimum.

\begin{proposition}[Fixed-potential rejection]\label{prop:rejection}
Fix a finite $\beta$ and propose independent words with law $q_\beta(X)=w(X)e^{\beta\cdot\chi(X)}/P(\beta)$. Among rejection procedures that accept the proposed word unchanged and have accepted law $w(X)/Z$ on $\F$, the greatest success probability is $Ze^{m_\F(\beta)}/P(\beta)$. Consequently the infimum expected proposal count over finite potentials is $U/Z$.
\end{proposition}
\begin{proof}
If success has probability $r$ and acceptance probability at $X$ is $a(X)$, exactness requires $q_\beta(X)a(X)=rw(X)/Z$ for $X\in\F$ and zero elsewhere. Since $a(X)\le1$, this forces $r\le Ze^{\beta\chi(X)}/P(\beta)$ for every feasible $X$. Equality is achieved by accepting a feasible $X$ with probability $e^{m_\F(\beta)-\beta\chi(X)}$. Independent trials then have mean $P(\beta)e^{-m_\F(\beta)}/Z$. Taking the infimum proves the claim, including a nonattained optimum.
\end{proof}

This statement restricts the proposal family and the acceptance mechanism. It supplies no lower bound for prefix-conditioned, stratified, or recursive exact samplers.

%% file: proofs/finite_optimization.tex
\section{Finite potentials and certified optimization}
\label{comp:finite}

For a nonempty family \(A\subseteq\Omega\), write
\[
 m_A(\beta)=\min_{X\in A}\beta\cdot\chi(X),\qquad
 G_A(\beta)=\log P(\beta)-m_A(\beta),\qquad
 U(A)=\inf_{\beta\in\mathbb R^N} e^{G_A(\beta)}.
\]
The source partition function is always evaluated on the original
complete quota product \(\Omega\), including when \(A\) is a relaxation
of the feasible family. All rational inputs and outputs are encoded in
binary.

Frozen chains may be removed from an oracle's dynamic program while
retaining their exact offsets. Let \(I\) be the remaining labels,
\(f\) the frozen incidence vector extended by zeros on \(I\), \(w_f\) its
selected activity product, and \(c_f(\beta)=\beta\cdot f\).
For \(A_I=\{X\cap I:X\in A\}\),
\[
 P(\beta)=w_f e^{c_f(\beta)}P_{\rm act}(\beta_I),\qquad
 m_A(\beta)=c_f(\beta)+m_{A_I}(\beta_I).
\]
Here \(P_{\rm act}\) is the partition function of the complete
quota product on \(I\).
Thus an active-word witness is extended by the frozen selections,
its lower support threshold is increased by \(c_f(\beta)\), and
its active-source capacity value is multiplied by \(w_f\).
These rational adjustments preserve the support contract below.
Equivalently, the optimizer may keep frozen potential coordinates
zero. The parameter \(N\) always counts the original labels.

\subsection{A finite box, including unattained infima}

Let \(A_\lambda\) be the total bit length of the numerators and
denominators of the activities. Then
\[
 |\Omega|\le 2^N,\qquad
 2^{-A_\lambda}\le w(X)\le 2^{A_\lambda},\qquad
 w(Y)/w(X)\le 2^{A_\lambda}.
\]
The last inequality follows by cancelling the factors for labels selected in both words.
A chain-constant change of \(\beta\) leaves \(G_A\) unchanged, since the
chain quota is fixed.

\begin{lemma}[Uniform clipping]\label{comp:clipping}
For every \(\beta\in\mathbb R^N\), \(H>0\), and nonempty
\(A\subseteq\Omega\), there is \(\bar\beta\in[0,2H]^N\) such that
\[
 \log P(\bar\beta)-\bar\beta\cdot x
 \le \log P(\beta)-\beta\cdot x+
             N2^{N+A_\lambda}e^{-H}
 \quad
 (x\in\operatorname{conv}\{\chi(X):X\in\Omega\}).
\]
Consequently the same additive error bounds
\(G_A(\bar\beta)-G_A(\beta)\).
\end{lemma}
\begin{proof}
Remove deterministic chains, retaining their mass as a global factor.
In each remaining chain let \(T_c\) consist of the \(g_c\) largest
potential coordinates. Subtract the smallest selected coordinate from
the chain. Coordinates in \(T_c\) are now nonnegative and all other
coordinates are nonpositive. This quota gauge changes no displayed
objective. Put \(T=\bigcup_c T_c\).

Throughout the following clipping, \(T\) maximizes the linear
potential over the source. If \(i\in T\) has current coordinate \(u\ge0\),
every source word omitting \(i\) has potential at most
\(\beta\cdot\chi(T)-u\). Comparison with the single denominator term
indexed by \(T\) gives
\[
 \Pr_\beta(i\notin X)\le 2^{N+A_\lambda}e^{-u}.
\]
For \(f_x(\beta)=\log P(\beta)-\beta\cdot x\), where \(0\le x_i\le1\),
\(\partial_i f_x\ge-\Pr_\beta(i\notin X)\).
Lowering a coordinate from \(u>H\) to \(H\) therefore increases
\(f_x\) by at most \(2^{N+A_\lambda}e^{-H}\).
If \(i\notin T\) has coordinate \(u\le0\), the analogous comparison
gives \(\Pr_\beta(i\in X)\le2^{N+A_\lambda}e^u\).
Since \(\partial_i f_x\le\Pr_\beta(i\in X)\), raising \(u<-H\) to
\(-H\) has the same error bound. These operations preserve the sign
pattern. Clip all coordinates successively and then add \(H\) to
each chain. Summing at most \(N\) errors proves the claim. The word
\(T\) is required to belong only to the source, not to \(A\).
Finally, \(G_A=\max_{X\in A} f_{\chi(X)}\).
\end{proof}

Given a rational internal tolerance \(0<\tau\le1\), set
\[
 H=N+A_\lambda+\left\lceil\log_2(8N/\tau)\right\rceil,\qquad R=2H.
\]
Because \(e^{-H}\le2^{-H}\),
\begin{equation}\label{comp:box}
 \min_{\beta\in[0,R]^N}G_A(\beta)
       \le \log U(A)+\tau/8.
\end{equation}
This conclusion follows by clipping a minimizing sequence and then
using compactness; a finite minimizer on the unrestricted space is
not assumed. Each subgradient of \(G_A\) is the difference between
the source marginal vector and a convex combination of the incidence
vectors of words attaining \(m_A\) at the given potential. Thus \(G_A\) is \(\sqrt N\)-Lipschitz in Euclidean norm
and \(N\)-Lipschitz in the maximum norm. In particular, rounding a
box minimizer to mesh \(\tau/(8N)\) produces a polynomial-bit
near-optimal potential.

The full-source quantities needed below admit polynomial-bit numerical
evaluation:
\[
 P(\beta)=\prod_c e_{g_c}
                  \bigl((\lambda_i e^{\beta_i})_{i\in C_c}\bigr),
 \qquad
 \partial_i\log P(\beta)=
 \frac{\lambda_i e^{\beta_i}
 e_{g_c-1}((\lambda_j e^{\beta_j})_{j\in C_c\setminus\{i\}})}
 {e_{g_c}((\lambda_j e^{\beta_j})_{j\in C_c})}.
\]
The elementary symmetric recurrence uses \(O(\sum_c n_cg_c)\)
operations, and prefix and suffix tables provide all marginals.
On the box, logarithms of nonzero intermediate values have magnitude
\(O(N+A_\lambda+NR)\), and every nonzero denominator is at least
\(2^{-A_\lambda}\). Range reduction and Taylor series give outward
rational intervals for exponentials; scaling to \([1,2]\) and the
atanh series give such intervals for logarithms. Positive-term
recurrences and these range bounds show that additive error \(\zeta\)
in \(\log P\) and in every marginal takes bit time and space
polynomial in \(N+A_\lambda+NR+\log(1/\zeta)\) and the binary
encoding lengths of \(\beta\) and the requested tolerance. For quota
zero or full quota the marginal is the corresponding deterministic
indicator.
Thus these computations do not presume real arithmetic or an oracle
for the true feasible mass.

\subsection{An approximate support oracle suffices}

\begin{lemma}[Finite-output optimization]\label{comp:finite-oracle}\label{comp:finite-optimizer}
Suppose that, for each rational \(\beta\in[0,R]^N\) and rational
\(\delta>0\), an algorithm returns \(X\in A\) and a rational \(t\)
such that
\begin{equation}\label{comp:support-contract}
 t\le m_A(\beta)\le\beta\cdot\chi(X)\le t+\delta.
\end{equation}
If its bit running time is polynomial in the instance and rational
query encoding lengths, \(R\), and \(1/\delta\), then there is a
deterministic FPTAS returning
rational \(\beta,t,V\) with
\[
 t\le m_A(\beta),\qquad
 U(A)\le P(\beta)e^{-t}\le V\le(1+\varepsilon)U(A).
\]
No exact optimization oracle for \(A\), and no attained optimum for
\(U(A)\), is required.
\end{lemma}
\begin{proof}
If the quota source has one word, then \(A\) consists of that
word. Return \(\beta=0\), \(t=0\), and its exact rational weight.
Otherwise use \(\tau=\varepsilon/4\), where \(0<\varepsilon\le1\), and the
box in \eqref{comp:box}. Put
\[
 \delta=\tau/16,\quad
 J=\left\lceil16NR/\tau\right\rceil,\quad
 T=J^2,\quad h=R/(2J),\quad\beta_1=0.
\]
At iteration \(a\) obtain \(X_a,t_a\) from
\eqref{comp:support-contract}. Compute a dyadic marginal vector
\(\widetilde p_a\), clipped to \([0,1]^N\), with
\[
 \|\widetilde p_a-\nabla\log P(\beta_a)\|_\infty
                         \le\tau/(16NR).
\]
Set
\[
 \widetilde g_a=\widetilde p_a-\chi(X_a),\qquad
 \beta_{a+1}=\operatorname{proj}_{[0,R]^N}
                            (\beta_a-h\widetilde g_a).
\]
For all \(z\), convexity of \(\log P\), feasibility of \(X_a\), and
\(\beta_a\cdot\chi(X_a)-m_A(\beta_a)\le\delta\) give
\[
 G_A(z)\ge G_A(\beta_a)+
 [\nabla\log P(\beta_a)-\chi(X_a)]\cdot(z-\beta_a)-\delta .
\]
For \(z\) in the box, the marginal error adds at most \(\tau/16\).
Thus \(\widetilde g_a\) is a \(\tau/8\)-subgradient on the box and
has norm at most \(\sqrt N\), hence certainly \(2\sqrt N\).

Let \(\beta^*\) minimize \(G_A\) on the box. The projection inequality,
expanded and summed over the iterations, gives
\[
 \frac1T\sum_{a=1}^T [G_A(\beta_a)-G_A(\beta^*)]
 \le \frac{NR^2}{2hT}+\frac{h(4N)}2+\frac{\tau}{8}
 =\frac{2NR}{J}+\frac{\tau}{8}\le\frac{\tau}{4}.
\]
Convexity gives the same bound for
\(\widehat\beta=T^{-1}\sum_a\beta_a\). Call the support oracle once
more at \(\widehat\beta\), obtaining \(\widehat t\). Combining the
box error, the iteration bound, and this last oracle error yields
\[
 \log P(\widehat\beta)-\widehat t
             \le\log U(A)+7\tau/16.
\]
Outward evaluation with additional logarithmic error at most
\(\varepsilon/4\) gives a rational \(V\) above the actual envelope
and below \(e^{\varepsilon/2}U(A)\le(1+\varepsilon)U(A)\).

For bit complexity, use one fixed dyadic denominator for all marginal
approximations. The fixed rational step, additions, and coordinate
clipping then keep all iterate denominators in one polynomial-bit
common denominator; averaging adds only \(\log T\) bits.
There are \(T+1\) support calls with \(T=O(N^2R^2/\tau^2)\). Source evaluation has
the bit bound proved above, and the assumed support oracle has
polynomial bit cost. Also
\(2^{-A_\lambda}\le U(A)\le2^{N+A_\lambda}\), so the final rational
upper value requires polynomially many bits.
\end{proof}

\subsection{The rank-one endpoint}

\begin{theorem}[Rank-one certified FPTAS]\label{comp:rank-one}
On an explicitly listed rational complete quota source with
commonly ordered covering rows of supplied exchange rank at most
one, and arbitrary positive rational activities, the all-cut capacity
has a deterministic certified FPTAS. Infeasibility is detected
exactly. The activity order is not required for this theorem.
\end{theorem}
\begin{proof}
First consider one covering row \(a\cdot\chi(X)\ge\theta\), allowing
signed scores and arbitrary quotas. For \(\beta\in[0,R]^N\) and
\(\delta>0\), put
\(\eta=\delta/N\) and \(k_i=\lfloor\beta_i/\eta\rfloor\).
Process labels chain by chain. At a fixed total integer cost and
current-chain quota count, retain the largest exact original row
score and a witness attaining it. Taking or skipping a label updates
the cost, quota count, and score; at a chain boundary enforce its
quota and reset the local count. Larger score dominates smaller
score at the same state for every possible suffix, including with
signed scores. Thus the minimum accepted integer cost is exactly
\[
 K^*=\min_{X\in\mathcal F}\sum_{i\in X} k_i.
\]
For its retained witness \(X\), take \(t=\eta K^*\). Since
\[
 \eta\sum_{i\in Y}k_i\le\beta\cdot\chi(Y)
              <\eta\sum_{i\in Y}k_i+\delta
 \quad(Y\in\Omega),
\]
the pair \(X,t\) satisfies \eqref{comp:support-contract}.
The largest cost index is at most \(N^2R/\delta\); scores are sums
of at most \(N\) input rationals and have polynomial bit length.
This is the required polynomial-time additive oracle. Feasibility
is checked by selecting the largest \(g_c\) row scores in each
chain. Lemma~\ref{comp:finite-oracle} proves the one-row claim.

Now remove frozen chains, retaining their exact contributions.
Rational elimination on within-chain score differences computes
the supplied rank. At rank zero all rows are constant on the source;
check each threshold, and if they all pass use \(\beta=0,t=0\).
At rank one choose a nonconstant supplied row \(a_*\).
Every other nonconstant row satisfies
\[
 a_j\cdot\chi(X)=c_j\,a_*\cdot\chi(X)+D_j
                       \quad (X\in\Omega)
\]
for computable rational \(c_j,D_j\). Indeed, proportional
within-chain differences leave a constant on each chain, whose
quota-weighted sum is \(D_j\). The common order forces \(c_j>0\):
a strict increase of \(a_*\) cannot be reversed by another
nondecreasing row. Constant rows are checked separately, and all
remaining inequalities are equivalent to
\[
 a_*\cdot\chi(X)\ge
       \max_{j\ {\rm nonconstant}}\frac{\theta_j-D_j}{c_j}.
\]
Apply the one-row algorithm. The maximum retains every original
inequality, including those linearly dependent on the chosen row.
\end{proof}

The cost-scaling dynamic program is a standard constrained-knapsack
technique; see Patel, Khan, and Louis~\cite{PatelKhanLouis2021}.
The finite-box argument is a quota-specific form of the bounded-dual
principle of Straszak and Vishnoi~\cite{StraszakVishnoi2019}.
Our use of these tools requires the exactly valid lower support
threshold in \eqref{comp:support-contract}, rather than a rounded
feasibility test.

%% file: proofs/count_sampling.tex
\section{Counting and exact sampling at fixed supplied rank}
\label{comp:count}

\begin{theorem}\label{comp:count-sample}
Consider a complete product of explicit fixed-quota chains.
Scores and positive rational activities are nondecreasing in a common
order on each chain. Let \(N\) be the number of original labels,
\(p\) the number of explicitly supplied covering rows, and \(q\)
their exchange rank. For every fixed \(q\):
\begin{enumerate}
\item the true feasible mass \(Z\) has a deterministic FPTAS returning
\(Z/(1+\varepsilon)\le\widehat Z\le Z\);
\item on every nonempty instance the law \(w(X)/Z\) can be sampled
exactly in expected polynomial bit time.
\end{enumerate}
The number \(p\) is part of the input. The sampler needs at most
\((1+N^2)^q\) proposals in expectation. The polynomial degree in
the counting and sampling bounds may depend on \(q\).
\end{theorem}

The algorithm uses the supplied representation. Low rank in a different,
unknown representation does not give the required input certificate.

Both algorithms use one rounded source. Exact sampling needs its suffix
masses and an acceptance-density bound; deterministic counting additionally
compresses true acceptance probabilities.

\subsection{The common rounded source}\label{comp:count-generator}
The top quota word \(T\) maximizes every row simultaneously: \(\F\)
is empty exactly when \(T\) fails a row. Record and remove frozen-chain
row contributions and activities; if no labels remain, evaluate the sole
word directly. Below \(N\) counts the remaining labels. The original
count only weakens our bounds.

\paragraph{Nonnegative losses for prefixes.}
In each nonfrozen chain choose its lowest top label as a pivot,
using the same label for every row. A selected bottom label is a
mismatch whose row charge is the pivot score minus its score.
An omitted top label is a mismatch whose charge is its score minus
the pivot score. All charges are nonnegative. Since the numbers of
selected bottom and omitted top labels agree in each chain, their
total charge is exactly the score lost relative to \(T\).

For row \(j\), let \(B_j\ge0\) be its score at \(T\) minus its
threshold. If \(B_j=0\), forbid every mismatch having positive
charge in that row. These are exact inclusion or exclusion pins.
For each \(B_j>0\), divide that row's charges by \(B_j\). Write \(A\)
for the resulting nonnegative matrix, with columns indexed by the
original labels, and let \(y\) be a mismatch vector. The remaining
tests are \(A_j y\le1\).
Its rank \(r\) is at most \(q\): every column before normalization is
a signed within-chain difference to the common pivot.
In particular the rank bound holds for incomplete prefixes, whose
quotas need not yet be satisfied. If \(r=0\), the quotas and pins
alone suffice, and exact weighted dynamic programming counts and
samples.

\paragraph{A rounded superset.}
Assume \(r>0\). Fix \(r\) independent columns of \(A\), and choose
\(r\) rows of maximum absolute determinant on those columns.
For fixed \(r\) this can be done by enumeration. Denote the selected
full rows by \(A_*\). Cramer's rule and replacement of any one
selected row show that
\[
 A_j=\alpha_j A_*,
 \qquad |\alpha_{j\ell}|\le1 \quad(1\le\ell\le r).
\]
This is the maximum-volume basis principle
\cite{AwerbuchKleinberg2004}; the determinant argument also proves
the coefficient bound here. This row basis controls the size of the
generator; the column basis below will control its excess mass.

Set \(H=8rN^2\). Round each entry of \(HA_*\) down to an integer,
and track the \(r\) cumulative rounded charges \(v\).
Reject a prefix if any coordinate exceeds \(H\).
At the final layer additionally require, for every positive-budget
row,
\[
 \alpha_j v/H\le 1+1/(8N).
\]
Together with the exact quotas and pins, these rules define a family
\(G\). The true and rounded basis totals differ by at most \(N/H\)
in each coordinate. The coefficients \(\alpha_j\) may be signed, but
\[
 |A_jy-\alpha_jv/H|\le\sum_{\ell=1}^r
 |\alpha_{j\ell}|N/H\le rN/H=1/(8N).
\]
Since the basis charges are
nonnegative, a truly feasible word survives all prefix caps.
Consequently
\begin{equation}\label{comp:G}
 \mathcal F\subseteq G,\qquad
 A_jy\le1+1/(4N)\quad(X\in G).
\end{equation}

Read labels chain by chain. A state is \(u=(i,h,v)\): the number
\(i\) of processed labels, the number \(h\) selected in the current
chain, and the cumulative rounded charges \(v\). Taking or skipping
the next label updates \(h\) and adds its rounded charge exactly
when that decision is a mismatch. Enforce pins and \(v\le H\)
coordinatewise. At the end of a chain require its exact quota and
reset \(h\) to zero. At the last layer enforce the reconstructed
row tests. This specifies a generator for \(G\), with
\(O((N+1)(H+1)^r)\) states at each layer.

Write \(\lambda_i=u_i/d_i\) in positive integers, using weight
\(u_i\) for taking label \(i\) and \(d_i\) for skipping it.
For an allowed edge \(e:u\to u'\), let \(a_e\) be this integer
weight. Set \(S(u)=1\) at accepting terminal states, zero at other
terminal states, and
\[
 S(u)=\sum_{e:u\to u'}a_e S(u'),\qquad
 t_e=\frac{a_eS(u')}{S(u)}\quad\text{when }S(u)>0.
\]
Delete states with \(S(u)=0\). For the initial state \(u_0\),
\(Z_G=S(u_0)/\prod_i d_i\), apart from the recorded frozen-chain
factor. The probabilities \(t_e\) define the common conditional
suffix law at \(u\): the product of the transition probabilities along a
live suffix telescopes to its integer weight divided by \(S(u)\),
independently of the prefix.

\subsection{Acceptance density and exact sampling}\label{comp:count-repair}
Fix \(X\in G\). Match its selected bottom labels bijectively to its
omitted top labels within each chain, using a deterministic rule.
Each matched pair defines a disjoint quota-preserving upgrade.
Its nonnegative improvement vector is the sum of its two charge
columns. The sum of all these vectors is the entire loss vector
\(Ay\).

Let their span have dimension \(d\le r\).
If \(d=0\), then \(X\) is already feasible.
Otherwise fix \(d\) independent row coordinates and select \(d\)
improvement columns of maximum absolute determinant.
Cramer's rule bounds every reconstruction coefficient in absolute
value by one. Because all these columns are nonnegative, every
improvement column \(v_e\) obeys the componentwise inequality
\[
 v_e\le\sum_{b\ {\rm selected}}v_b .
\]
There are at most \(N\) pairs. Thus performing the selected basis
upgrades removes at least \(1/N\) of the total charge in every row.
The remaining charge is at most
\((1-1/N)(1+1/(4N))\le1\), by \eqref{comp:G}.
Zero-budget pins survive because an upgrade only erases mismatches.
The repaired word belongs to \(\mathcal F\), and its weight is at
least \(w(X)\), by the common activity order.

Fix one deterministic repair choice for every \(X\in G\).
Given a repaired word and an ordered inverse-swap list, its preimage
is uniquely determined. A list of at most \(r\) swaps, padded with
no-ops, has at most
\[
 K_0=(1+N^2)^r
\]
possibilities. Summing weights over the fibers gives
\begin{equation}\label{comp:accept-density}
 Z\le Z_G\le K_0Z.
\end{equation}
The generator samples exactly from \(w(X)/Z_G\) on \(G\).
Rejecting a proposal precisely when it fails an original row
therefore samples exactly from \(w(X)/Z\), with at most \(K_0\)
expected proposals. Rational transitions are sampled by fair-bit rejection for uniform
integer draws, with fresh bits and polynomial expected bit cost.
The sampler never computes the repair map.

\subsection{Deterministic compression of acceptance probabilities}\label{comp:count-compression}
\paragraph{Backward CDF tables.}
Clear denominators in each positive-budget row, obtaining integer
mismatch charges and budget \(\bar B_j\). This scaling preserves rank;
it changes neither the generator nor its transition probabilities.
For requested error \(0<\varepsilon\le1\) (replace a larger
\(\varepsilon\) by one), put
\(\delta=\varepsilon/(2+\varepsilon)\). Choose an integer
\[
 M=\left\lceil 4(p+1)(N+1)^2K_0/\delta \right\rceil.
\]
Let \(c_j(e)\) be the exact integer mismatch charge of edge \(e\)
in row \(j\), after that row's denominator clearing. At every live
state \(u\) in layer \(i\), let \(F_{ju}(b)\) be the conditional
probability that the remaining suffix charge in row \(j\) is at
most the integer \(b\). At a terminal state set
\(\widehat F_{ju}(b)=\mathbf1_{\{b\ge0\}}\); recursively set
\begin{equation}\label{comp:cdf-recursion}
 \widehat F_{ju}(b)=\frac1M\left\lfloor
 M\sum_{e:u\to u'}t_e\widehat F_{ju'}(b-c_j(e))
 \right\rfloor.
\end{equation}
The exact CDFs satisfy the unrounded recurrence. Induction adds at
most \(1/M\) error per layer, giving
\[
 0\le F_{ju}(b)-\widehat F_{ju}(b)\le (N-i)/M
 \quad\text{for every integer }b.
\]
Each approximate CDF is nondecreasing and has at most \(M+1\)
distinct values. Merge the two shifted breakpoint lists in
\eqref{comp:cdf-recursion}, evaluate the rational sum on their
constant intervals, round, and coalesce equal adjacent values.
Zero-probability edges are omitted. There is no iteration over the
numerical score range; negative remaining budgets are valid queries.

\begin{lemma}[Prefix compression invariant]\label{comp:prefix-invariant}
Consider a layered generator of length \(N\), with a common
conditional suffix law \(Q_u\) for all prefixes reaching a live state
\(u\). For each row \(j\), all such prefixes use the same suffix charge
\(C_{ju}(s)\). A suffix \(s\) completes prefix \(\pi\) feasibly exactly when
\(C_{ju}(s)\le b_j(\pi)\) in every row. There are at most \(p\) rows.
At layer \(i\), suppose the CDF of each \(C_{ju}\) under \(Q_u\) is
approximated uniformly within \(e_i\). Give a prefix the key consisting of its
state and these approximate CDF values at its remaining budgets.
Extend representatives using the conditional generator transitions;
at each key retain any actual candidate prefix and add the propagated
probabilities. If \(a\) is the initial acceptance probability and
\(\widehat a\) is the total terminal probability of accepting
representatives, then
\[
 |\widehat a-a|\le 2p\sum_{i=1}^N e_i.
\]
\end{lemma}
\begin{proof}
For two prefixes at one key, the suffix law is the same. In one row
their remaining-budget events are nested, so their symmetric
difference has probability at most \(2e_i\). The union bound gives
\(|h(\pi)-h(\pi')|\le2pe_i\), where \(h(\pi)\) is the true
conditional probability of accepting after prefix \(\pi\).
This argument does not assume independence between rows.

After the merges at layer \(i\), set
\(\Phi_i=\sum_C\omega_C h(\pi_C)\), where \(\omega_C\) is the
propagated probability of a key and \(\pi_C\) its representative.
Extension preserves this sum exactly by conditioning on the next
transition. A merge changes it by at most \(2pe_i\) times the
merged probability. The total propagated probability remains one,
whence
\[
 |\Phi_i-\Phi_{i-1}|\le2pe_i,\qquad
 \Phi_0=a,\qquad\Phi_N=\widehat a.
\]
Telescoping proves the claim. The propagated probability belongs to
the whole key; it need not equal the probability of its representative.
\end{proof}

\paragraph{Simultaneous keys.}
For an actual prefix \(x\) at layer \(i\), let \(s_j(x)\) be its
exact integer charge in positive-budget row \(j\). Keep its generator
state \(u\) and the key
\[
 \kappa_u(x)=
 \bigl(\widehat F_{ju}(\bar B_j-s_j(x))\bigr)_{j:B_j>0}.
\]
Zero-budget pins are already enforced by the generator.
Lemma~\ref{comp:prefix-invariant} applies with
\(e_i=(N-i)/M\). In particular, two prefixes at the same state and key have true
conditional acceptance probabilities differing by at most
\(2p(N-i)/M\), including keys on exact breakpoint boundaries.

The forward pass starts with the empty prefix and unit mass.
At each layer the update is as follows.
\begin{enumerate}
\item For each occupied state and key, extend its actual representative
      along every live successor transition and multiply its propagated
      probability by that transition probability.
\item Evaluate each candidate's new key at its exact remaining
      row budgets. Group candidates by successor state and key.
\item At each occupied successor state and key, keep the
      lexicographically first actual candidate prefix, using
      \(0<1\), and the sum of the candidates' propagated probabilities.
\end{enumerate}
At the terminal layer add the propagated probabilities of
representatives that pass all original rows.

The lemma's potential invariant gives, after layer \(i\),
\begin{equation}\label{comp:forward-invariant}
 \left|\Phi_i-\frac Z{Z_G}\right|
 \le\frac{2p}{M}\sum_{h=1}^i(N-h)\le\frac{2pNi}{M}.
\end{equation}
At termination, \(\Phi_N\) is the acceptance estimate and has error
at most \(2pN^2/M\). Multiplication by \(Z_G\) and
\eqref{comp:accept-density} give relative error at most
\[
 K_0\,2pN^2/M\le\delta/2.
\]

For the resulting mass estimate \(E\), use the weaker bound
\(|E-Z|\le\delta Z\) and return \(\widehat Z=E/(1+\delta)\):
\[
 \frac{Z}{1+\varepsilon}
 =\frac{1-\delta}{1+\delta}Z
 \le\widehat Z\le Z.
\]

\subsection{State counts and bit complexity}\label{comp:count-size}
\paragraph{Reachable keys.}
At a fixed layer and state, true prefix-charge vectors lie in a
linear space of dimension at most \(r\), since they are partial
sums of the same row-scaled charge columns. Each row CDF contributes at most
\(M\) affine hyperplanes \(s_j=\bar B_j-b\), where \(b\) ranges
over its stored breakpoints. Including lower-dimensional
faces to handle exact ties, an arrangement of \(m\le pM\)
hyperplanes in dimension \(r\) has \((1+m)^{O(r)}\) faces.
Indeed, inserting hyperplanes gives at most
\(\sum_{k=0}^r\binom{m}{k}\) full-dimensional regions. Each face has
an affine hull specified by at most \(r\) independent hyperplanes;
apply the same region bound in that hull and sum over the choices.
The key is constant on each face, using the CDF convention
``charge at most budget'' on equality faces. Hyperplanes that vanish
identically on the charge space impose constant tests.
Thus each state has at most \((1+pM)^{O(r)}\) keys. Only reachable
keys are generated; no arrangement is constructed. With
\(\mathsf S=O((N+1)^2(H+1)^r)\) generator states over all layers,
the backward tables store \(O(\mathsf S p(M+1))\) entries and the
forward pass visits at most \(\mathsf S(1+pM)^{O(r)}\) representatives.

\paragraph{Bit complexity.}
For row \(j\), multiply the unnormalized pivot losses and budget by the
product of the denominators of its input scores and threshold. This gives
integer charges and budget \(\bar B_j\) with polynomial bit length.
Each breakpoint sums at most \(N\) such charges. Clearing denominators
and applying Hadamard's bound also bounds the bit lengths of the
determinants in each basis ratio by a polynomial in the input length.
Conditional transition probabilities
are ratios of integer suffix masses. New CDF values are rounded
back to the fixed denominator \(M\), so their denominators do
not accumulate.

The forward computation can use unnormalized integer prefix
masses. If \(S(u)\) is the integer suffix mass at state \(u\),
an aggregate prefix mass \(W_C\) represents probability
\(W_C S(u)/S(u_0)\), independently of its chosen representative.
Extending by integer transition weight \(a\) sends \(W_C\) to \(W_Ca\);
the suffix-mass factors give exactly the required conditional
transition probabilities. Merges occur at the same state, so this
interpretation survives merging. Each original decision history follows a
unique path through the compressed computation. At a merge its generator
state is preserved, so each next edge is legal for every history carried
by the record. Route all histories taking that edge to the key computed
from the extended representative. Their own exact budgets may give other
keys. The two decisions partition the extended histories, and merging
records sums disjoint history sets. Thus no history is counted twice. Every stored integer mass is bounded
by \(\prod_i(u_i+d_i)\), and has polynomial bit length.
At the terminal layer the common activity denominator and recorded
frozen-chain factor convert the accepted mass into \(E\).
For fixed \(q\), all state counts and all operations are polynomial
in the input length and \(1/\varepsilon\).
\begin{proof}[Proof of Theorem~\ref{comp:count-sample}]
Preprocessing handles emptiness, frozen chains, and rank zero.
For positive rank, \eqref{comp:accept-density} gives exact sampling,
and \eqref{comp:forward-invariant} gives the one-sided count after
rescaling. Section~\ref{comp:count-size} supplies polynomial bit bounds;
\(r\le q\) gives the stated proposal bound.
\end{proof}

The repair and inverse-image estimate are related to the rounding
principle used by Dyer~\cite{Dyer2003}; deterministic compression of
knapsack distributions has precedents in
\cite{GopalanEtAl2011,GopalanKlivansMeka2012}.
Here the simultaneous rank bound controls all explicitly supplied
rows, and the conditional suffix law comes from a weighted complete
quota product. Commonly ordered activities are used exactly in
the comparison \(w(X)\le w(\operatorname{repair}(X))\);
without them \eqref{comp:accept-density} need not hold.

%% file: proofs/capacity_upper.tex
\section{Certified capacity approximation at fixed supplied rank}
\label{comp:upper}

Let \(s=1\) when every nonfrozen chain selects one or omits one label,
and \(s=2\) otherwise. An \emph{upgrade} replaces a selected label
by a later unselected label in the same common order.

\begin{theorem}\label{comp:rank-upper}
Fix \(q\ge1\). On the rational commonly ordered inputs of
Theorem~\ref{comp:count-sample} with supplied exchange rank at most
\(q\) and nonempty feasible family, a polynomial-bit-time algorithm
returns rational \(\beta,t,V\) such that
\[
 t\le m_{\mathcal F}(\beta),\qquad
 U(\mathcal F)\le P(\beta)e^{-t}\le V
               \le128N^{s(q-1)}U(\mathcal F).
\]
There is consequently a two-sided value approximation of factor
\(O(N^{s(q-1)/2})\).
At \(q=1\), Theorem~\ref{comp:rank-one} gives the stronger
\(1+\varepsilon\) guarantee. At \(q=2\), the powers \(s\) for
certified or bare potentials and \(s/2\) for two-sided values are
optimal, in the sense of
Theorems~\ref{comp:bare-hardness} and \ref{comp:value-hardness}.
The matching bounds determine the exponents of \(N\).
\end{theorem}

\begin{lemma}[Capacity stability under upgrades]\label{comp:upgrade}
Let \(\varnothing\ne A\subseteq D\subseteq\Omega\).
Suppose that every word of \(D\) reaches \(A\) by at most \(k\)
upgrades. For positive commonly ordered activities and \(N\ge2\),
\[
 U(A)\le U(D)\le e^{1+e}N^{sk}U(A).
\]
\end{lemma}
\begin{proof}
Fix a finite potential \(\beta\). In a nonfrozen chain of size \(n\)
and quota \(g\), use effective weights \(u_i=\lambda_i e^{\beta_i}\).
We modify them to have maximum-to-minimum ratio at most \(N^s\),
while increasing every word objective
\(\log P(\beta)-\beta\cdot\chi(X)\) by at most \(1+e\) over
all chains.

Put \(T=e_g(u)/e_{g-1}(u)>0\) and \(a=T/N\).
First replace \(u_i\) by \(v_i=\max(u_i,a)\).
Since \(v_i\le u_i+a\),
\[
 e_j(u+a\mathbf1)=
 \sum_{\ell=0}^j\binom{n-j+\ell}{\ell}a^\ell e_{j-\ell}(u).
\]
The elementary-symmetric ratios \(e_j/e_{j-1}\) are nonincreasing
by Newton's inequalities. Therefore
\(e_{g-\ell}(u)\le e_g(u)T^{-\ell}\) and
\(e_{g+1-\ell}(u)\le e_g(u)T^{1-\ell}\).
Using \(\binom{n-j+\ell}{\ell}\le n^\ell/\ell!\), we obtain
\[
 e_g(v)/e_g(u)\le e^{n/N},
 \qquad
 e_{g+1}(v)/e_g(v)\le T e^{n/N}.
\]
The increase of a word objective is at most the first logarithmic
ratio: its selected effective-weight denominator only increases.

Next cap \(v_i\) at \(TN\). In omitted-label coordinates this is
flooring \(v_i^{-1}\) at \(1/(TN)\), with quota \(n-g\), since
\[
 \frac{e_g(v)}{\prod_{i\in X}v_i}
 =\frac{e_{n-g}(v^{-1})}{\prod_{i\notin X}v_i^{-1}}.
\]
The relevant consecutive-sum ratio is
\[
 \frac{e_{n-g-1}(v^{-1})}{e_{n-g}(v^{-1})}
          =\frac{e_{g+1}(v)}{e_g(v)}
          \le T e^{n/N}.
\]
Applying the same expansion bounds the second logarithmic cost by
\((n/N)e^{n/N}\). The final weights lie in \([T/N,TN]\).
Summing over chains costs at most \(1+e\), because
\(\sum_c n_c\le N\) and \(n_c/N\le1\).

If \(g=1\), only the first floor is needed: \(T=e_1(u)\) already
bounds each \(u_i\) above. If \(g=n-1\), apply this one-floor
argument directly to the omitted weights. When all chains have
one of these quotas, the ratio is at most \(N\) and the total
logarithmic cost is at most one.

Let \(\beta'_i=\log(u'_i/\lambda_i)\) be the modified potential.
The original factor \(w(X)\) cancels in the wordwise comparison
above. An upgrade \(i\to j\) increases \(\beta'\cdot\chi(X)\) by
\[
 \beta'_j-\beta'_i
 =\log(u'_j/u'_i)-\log(\lambda_j/\lambda_i)\le s\log N.
\]
Hence \(m_D(\beta')\ge m_A(\beta')-sk\log N\).
Taking maxima of the wordwise objective comparison over \(A\)
gives \(G_A(\beta')\le G_A(\beta)+1+e\), so
\[
 G_D(\beta')\le G_A(\beta)+1+e+sk\log N.
\]
Take the infimum over finite \(\beta\). This proves the upper
bound even at an unattained infimum. The lower bound follows
from inclusion.
\end{proof}

\begin{proof}[Proof of Theorem~\ref{comp:rank-upper}]
Test the top word as in Theorem~\ref{comp:count-sample}.
If the source has one word, return its exact weight with
\(\beta=0,t=0\). Otherwise remove frozen chains from the
exchange-coordinate construction, retaining their exact row
contributions and activity product. Restore their forced choices
in every witness.
Throughout, \(N\) still counts all original labels.
Using the same common pivots, let \(a_i\in\mathbb Q_{\ge0}^p\)
be the unnormalized mismatch-charge column of label \(i\), and
let \(b\ge0\) be the exact row-loss budget vector.
The columns have rank at most \(q\).

\paragraph{Step 1: fix the relaxed family before querying a potential.}
Enumerate all independent tuples
\(B=(b_1,\ldots,b_r)\) of distinct labelled charge columns,
where \(0\le r\le q\), fixing their order by label index.
In the chart indexed by \(B\), require its labels to be
mismatches. Permit another mismatch label \(i\) only if
\(a_i=Bc_i\) with \(\|c_i\|_\infty\le1\).
For \(r=0\), permit only zero charge columns.
For a word of positive active rank \(r\), fix \(r\) row coordinates
on which its active span projects injectively. Choose an active basis
\(B\) of maximum absolute projected determinant. Replacing one column
by any active column cannot increase this maximum, so Cramer's rule
gives \(\|c_i\|_\infty\le1\). A word of active rank zero uses the
zero-column chart.

For \(r>0\), round the first \(r-1\) coefficients of each \(c_i\)
down to multiples of \(1/(2N)\), including negative coefficients,
and retain its last coefficient exactly.
If \(z\) and \(\widehat z\) are the true and rounded coefficient
sums of a word, then
\[
 0\le z_j-\widehat z_j\le1/2\quad(j<r),
 \qquad z_r=\widehat z_r .
\]
In this chart accept precisely the words with \(B\widehat z\le b\).
Let \(D\) be the union of these accepted families, including the
zero-column chart. This family is fixed before any potential query
and does not depend on the oracle tolerance.
Since each basis column is nonnegative,
every feasible word passes in any chart containing it, and
\(\mathcal F\subseteq D\).

\paragraph{Step 2: bound its repair radius by \(q-1\).}
In an accepted word repair the pinned mismatch labels belonging
to the first \(r-1\) basis columns.
On a chain with \(u\) pinned selected bottom labels and \(v\)
pinned omitted top labels, extend both lists using other
mismatches to length \(\max(u,v)\), then pair them.
The fixed quota provides enough mismatches of both kinds.
These upgrades are disjoint; over all chains their number is at
most \(r-1\). They remove componentwise at least
\(\sum_{j<r}b_j\), while the rounding error is at most
\(\frac12\sum_{j<r}b_j\). Thus the repaired true loss is at most
\(B\widehat z\le b\). It belongs to \(\mathcal F\).
For \(r\le1\), the chart test is already exact.
Lemma~\ref{comp:upgrade} yields
\begin{equation}\label{comp:D-transfer}
 U(\mathcal F)\le U(D)
       \le e^{1+e}N^{s(q-1)}U(\mathcal F).
\end{equation}

\paragraph{Step 3: optimize rounded cost on that fixed family.}
Let \(I\) be the nonfrozen labels and \(c_f(\beta)\) the exact
potential of the forced selected labels, as in
Appendix~\ref{comp:finite}. The dynamic program processes only \(I\).
For a rational box potential \(\beta\) and error \(\delta\), round
the selected-label costs down on mesh \(\eta=\delta/N\), writing
\(k_i=\lfloor\beta_i/\eta\rfloor\) for \(i\in I\).
Here is the full state and recovery rule for a chart of rank \(r>0\).
For a label forbidden to be a mismatch, use \(c_i=0\) only as a
placeholder in the recurrence; its mismatch bit is forced to zero.
After a given number of labels have been read in chain order,
a state is
\[
 (h,K,v_1,\ldots,v_{r-1}),
\]
where \(h\) is the selected count in the current chain,
\(K\) is the total selected integer cost, and
\(v_j=\sum_{i\ {\rm mismatched}}\lfloor2Nc_{ij}\rfloor\).
The record stores the smallest exact sum
\(z_r=\sum_{i\ {\rm mismatched}}c_{ir}\) at this state and
one actual prefix attaining it. It starts with all indices zero.
For each allowed take/skip decision \(x\in\{0,1\}\), add
\(x\) to \(h\) and \(xk_i\) to \(K\). Put
\(e_i=x\) for a bottom label and \(e_i=1-x\) for a top label;
then add \(e_i\lfloor2Nc_{ij}\rfloor\) to \(v_j\) and
\(e_ic_{ir}\) to \(z_r\).
Pinned basis labels require \(e_i=1\); labels not permitted in
the chart require \(e_i=0\). At each chain boundary
enforce its exact quota and reset \(h\).

The invariant is: at each layer and state, the record is an actual
chart prefix and its exact last coordinate is the minimum over all
chart prefixes with those indices. When records meet, retain the least
\(z_r\), with a fixed tie rule and its witness. The two prefixes have the same quota count and face the same future
pins. Appending any suffix respecting those constraints therefore adds
the same rounded indices and exact last-coordinate increment to both.
The smaller last coordinate can only improve each terminal row test,
since the original basis column \(b_r\) is nonnegative. This proves
statewise dominance by induction, including negative coefficient
increments. No intermediate row test is used: partial coefficient
sums need not increase. For \(r=0\), omit all
coefficient indices, permit only zero-charge mismatches, and keep
any witness at each quota--cost state.

At the terminal layer a rank-zero chart passes automatically;
for \(r>0\) the acceptance test is exactly
\[
 B\bigl(v_1/(2N),\ldots,v_{r-1}/(2N),z_r\bigr)\le b.
\]
These tests use the rounded chart loss in every original row.
An accepted witness belongs to
\(D\) and need not belong to \(\mathcal F\).
The invariant shows that the least accepted cost over all charts is
\[
 K^*=\min_{Y\in D}\sum_{i\in Y\cap I}k_i.
\]
Indeed, every chart word has a retained prefix with no larger last
coordinate at each layer, while every retained record is an
actual chart prefix. Restore the frozen selections to recover the
full original-label witness \(X\), and set
\(t=c_f(\beta)+\eta K^*\). For every source word \(Y\),
\[
 c_f(\beta)+\eta\sum_{i\in Y\cap I}k_i
 \le\beta\cdot\chi(Y)
 \le c_f(\beta)+\eta\sum_{i\in Y\cap I}k_i+\delta.
\]
Consequently,
\[
 t\le m_D(\beta)\le\beta\cdot\chi(X)\le t+\delta .
\]
The witness certifies the support contract on \(D\). Its threshold
also certifies a lower bound for the original family, since
\[
 t\le m_D(\beta)\le m_{\mathcal F}(\beta),
 \qquad \mathcal F\subseteq D.
\]
Repair is used to compare capacities; the algorithm need not repair
\(X\) to return this threshold. Charts may overlap because the oracle
minimizes over their union rather than summing their masses.

There are \(N^{O(q)}\) charts. Every rounded coefficient sum is
in \([-2N^2,2N^2]\) in integer units, and the potential-cost range
is at most \(N^2R/\delta\).
Determinant ratios defining the exact coefficients and their
partial sums have polynomial bit length. Thus this oracle runs
in polynomial bit time at every fixed \(q\), without iterating
over a binary-encoded score range.

\paragraph{Step 4: recover the original-family certificate.}
Apply Lemma~\ref{comp:finite-oracle} to the family \(D\) with
relative tolerance one. Its proof uses only the displayed
support contract and the original-source partition function.
It returns a pair valid on \(D\), hence on \(\mathcal F\), and
\[
 U(D)\le P(\beta)e^{-t}\le V\le2U(D).
\]
Combine this with \eqref{comp:D-transfer} and
\(2e^{1+e}<128\).
For a two-sided value, let
\(L=\lceil\sqrt{128N^{s(q-1)}}\rceil\) and return \(V/L\).
Because \(L^2\ge128N^{s(q-1)}\), this number belongs to
\([U(\mathcal F)/L,L\,U(\mathcal F)]\).
The square-root ceiling is computed by integer arithmetic.
\end{proof}

%% file: proofs/hardness_gadget.tex
\section{A rank-two labelled multiplicity construction}
\label{comp:gadget-section}

The two hardness reductions use the same source and feasible family.
We first encode that family with two increasing rows, then record the
original-label means that the decoders will use. The source is never
restricted to the words that survive the encoding.

Fix \(s\in\{1,2\}\), \(K\ge1\), and an integer \(b\ge1\).
A query contains \(K\) lists \((w_{ri})_{i=1}^a\) of nonnegative
integers, indexed by \(0\le r<K\), with common even length \(a\ge4\).
Set
\[
 c=2^{b\lceil\log_2(a+1)\rceil},\qquad
 g=4aK,\qquad V=c^s.
\]
For each \((i,r,\eta,\sigma)\), where \(\eta\in\{0,1\}\) and
\(\sigma\in\{-1,1\}\), create one group. For \(s=1\), a group
consists of \(c\) binary quota-one chains; for \(s=2\), it is
one quota-\(c\) chain with \(c\) low and \(c\) high labels.
A defect is a selected low label. In the half-quota case the same
number of high labels is omitted. Thus a group has
\[
 p_d=\binom cd^s
 \quad\hbox{labelled words with \(d\) defects},\qquad
 P_c(z)=\sum_{d=0}^c p_dz^d.
\]
All activities are one. There are \(N=2cg=8aKc\) original labels.

\begin{lemma}\label{comp:gadget}
Two common-order integer covering rows define on this source exactly:
\begin{enumerate}
\item the \emph{bulk}: at most \(a/2\) positive-sign defects and at
most \(a/2\) negative-sign defects, with strict inequality for at
least one sign;
\item the \emph{corner}: one defect per item, all defects of a single
color \(r\), choices \(\eta_i\) satisfying
\(2\sum_i\eta_i w_{ri}=\sum_iw_{ri}\), and balanced defect signs.
\end{enumerate}
The scores can be made strictly increasing on every chain.
Every query, in both decision branches, has minimum common-order
representation rank exactly two on the same original quota source,
even if an equivalent representation is allowed to reorder each chain.
For fixed \(K,b\), the construction has polynomial binary length.
\end{lemma}
\begin{proof}
\textbf{The radix equality encodes only the corner.}
Put \(W_r=\sum_iw_{ri}\) and \(W_*=\max_rW_r\). Assign to each
group a nonnegative integer \(v=v_{ir\eta\sigma}\). Its \(2a\)
digits, and the digits of the target integer \(D\), are as follows:
\[
\begin{array}{c|c|c}
 \text{digit} & \text{contribution of group }(i,r,\eta,\sigma)
                 & \text{target}\\ \hline
 h\in[a] & \mathbf1_{\{i=h\}} & 1\\
 j=2,\ldots,a
   & (K-1-r)\mathbf1_{\{i=1\}}+r\mathbf1_{\{i=j\}} & K-1\\
 \text{last}
   & 2\eta w_{ri}+\mathbf1_{\{i=1\}}(W_*-W_r) & W_*
\end{array}
\]
Use radix
\[
 T=1+(cg+1)(3W_*+K+1).
\]
A word has at most \(c\) defects in each of the \(g\) groups.
Hence every digit sum in \(\sum vd\) is smaller than \(T\),
even for a rejected word. There are no carries.
The first \(a\) digits of \(\sum vd=D\) force one defect per
item. The next \(a-1\) force every item's color to equal item
\(1\)'s color. The final digit then says
\(2\sum_i\eta_iw_{ri}=W_r\). This equivalence includes zero
weights and repeated lists. The sign \(\sigma\) is deliberately
absent from these digits.

\paragraph{Two rows separate the bulk from the corner.}
Set
\[
 A=D+c\sum v,\qquad C=4(A+1),\qquad
 J=(a+1)C,\qquad B=(a+2)C.
\]
Give every low label score zero and every high label score
\(B\pm(v+J\sigma)\) in the two rows. Both increments are positive,
because \(|v|\le A<C=B-J\). Set each threshold to its all-high
score minus \( (aB\pm D) \), with matching signs.

For a word define its defect deficit, sign imbalance, and radix
error by
\[
 t=a-\sum d,\qquad S=\sum\sigma d,\qquad E=\sum vd-D.
\]
The respective row slacks are \(tB\mp(JS+E)\), so their minimum is
\[
 tB-|JS+E|,\qquad |E|\le A.
\]
If \(t<0\), their sum is negative. If \(1\le t\le a\), then
\[
\begin{array}{ll}
 |S|\le t:
   & tB-|JS+E|\ge tC-A>0,\\
 |S|\ge t+1:
   & tB-|JS+E|\le tC-J+A<0.
\end{array}
\]
Writing \(d_+\) and \(d_-\) for the two sign totals,
\(|S|\le t\) is equivalent to
\(\max(d_+,d_-)\le a/2\). Together with \(t\ge1\), this is
exactly the bulk condition. At \(t=0\), feasibility is
\(JS+E=0\). Since \(A<J\) and \(S\) is integral, this means
\(S=E=0\): balanced signs and the radix equality, exactly the
corner condition.

\paragraph{Strict scores preserve the family.}
Multiply both rows and thresholds by \(N^2+1\), then add the
zero-based within-chain position to every label score in both
rows. Any word receives an added total in \([0,N^2)\).
An old integral negative slack remains negative, and an old
nonnegative slack remains nonnegative. Within every chain the
new scores are strictly increasing.

\paragraph{A four-word trade rules out rank one.}
Choose one color and \(a-1\) distinct items. For each chosen
item fix an actual one-defect choice in a positive group and
one in a negative group. Form two words using all the positive
choices and all the negative choices, respectively; all other
groups are at their top words. Both words fail. Exchange their
choices on exactly \(a/2\) items. The resulting two words have
\(t=1\) and \(|S|=1\), so both are in the bulk, while the sum
of their original incidence vectors is unchanged. One signed
additive threshold therefore cannot accept both new words and
reject both old words.

Any exact common-order representation of rank one collapses to
one signed threshold: modulo chain constants its nonconstant
rows are proportional, and their proportionality coefficients
are positive because a common order forbids reversal of a
strict increase. Taking the maximum normalized threshold gives
the single inequality. This argument applies to every reselected
common order. Rank zero is impossible because the family is
nonempty and proper. The supplied two rows thus give minimum
representation rank exactly two.

Finally, \(c\le2^b(a+1)^b\). The radix has
\(O(\log(cg)+\log(1+W_*+K))\) bits; its \(2a\) positions, the
subsequent sums and products, and strictification all have
polynomial bit length. Every original label and both scores
are listed explicitly.
\end{proof}

\subsection{Symmetric group laws and original-label means}

At odds \(z>0\), assign each labelled group word with \(d\)
defects probability \(z^d/P_c(z)\). Its mean defect is
\[
 m(z)=zP_c'(z)/P_c(z).
\]
It increases strictly with \(\log z\), since the derivative is
the positive variance, and ranges from zero to \(c\).
The coefficient inequalities
\(d\,p_d\le Vp_{d-1}\) and \(p_1=V\) give
\begin{equation}\label{comp:odds-bounds}
 \frac{Vz}{1+Vz}\le m(z)\le Vz.
\end{equation}
For the upper bound, sum the coefficient inequality after
multiplying by \(z^d/P_c(z)\). For the lower bound, use
\(m(z)\ge\Pr(d\ge1)=1-P_c(z)^{-1}\) and \(P_c(z)\ge1+Vz\).
The entropy as a function of the mean is
\[
 H_c(m)=\log P_c(z)-m\log z
 \quad\text{when }m(z)=m,\qquad H_c'(m)=-\log z.
\]
Every low label has inclusion probability \(m/c\), and every
high label has inclusion probability \(1-m/c\). For binary
chains the symmetry permutes slots; for a half-quota chain it
permutes the low and high pools separately.

\begin{lemma}[Color means]\label{comp:color-means}
Define the rational vector \(\mu_r\) by giving the groups of
color \(r\) mean defect \(1/4\), uniformly over their original
labels, and making all other groups all high. If list \(r\)
is soluble, \(\mu_r\) is the mean of a feasible corner law.
If \(Q_z\) is the product group law with \(m(z)=1/(4K)\), then
\[
 \frac1K\sum_{r=0}^{K-1}\mu_r=\mathbb E_{Q_z}\chi .
\]
\end{lemma}
\begin{proof}
Fix a partition solution of list \(r\). Mix it fairly with its
complement, independently choose a uniform balanced sign vector,
and choose each one-defect labelled pattern uniformly. Every
word is in the corner. For each item, the side and sign are
independent fair bits, so each of its four groups has a defect
with probability \(1/4\). Uniform patterns give every low label
in that group inclusion probability \(1/(4c)\), and every high
label omission probability \(1/(4c)\). All other groups are
all high, proving the first assertion. Averaging these explicit
vectors over colors gives low-label inclusion \(1/(4Kc)\) and
high-label inclusion \(1-1/(4Kc)\), exactly the marginals of
\(Q_z\). The vectors remain defined for insoluble lists;
feasibility is asserted only for soluble colors.
\end{proof}

For later use, the finite Gibbs inequality on this unit-weight
source says, for every law \(Q\) and every signed potential,
\[
 \log P(\beta)\ge H(Q)+\beta\cdot\mathbb E_Q\chi .
\]
If \(\mathbb E_Q\chi\) is the mean of a feasible law, then
\(m_{\mathcal F}(\beta)\le\beta\cdot\mathbb E_Q\chi\), so
\(\log U(\mathcal F)\ge H(Q)\).
These are certificates at every original label. In particular
they apply to returned potentials that break all group symmetries.

%% file: proofs/bare_hardness.tex
\section{Bare-potential hardness by deterministic list pruning}
\label{comp:bare-section}

The decoder compares rational coordinates of the returned potential.
Its analysis uses a real Gibbs law, but the algorithm evaluates
neither that law nor the potential's unknown tight threshold.

\begin{theorem}\label{comp:bare-hardness}
For every \(s\in\{1,2\}\) and every fixed \(0<\gamma<s\),
there is a deterministic polynomial-time Turing reduction from
\textnormal{\textsc{Partition}} to returning a finite rational potential
\(\beta\) satisfying
\[
 P(\beta)e^{-m_{\mathcal F}(\beta)}
                       \le N^\gamma U(\mathcal F).
\]
Every query has unit activities, two strictly increasing integer
covering rows, and minimum common-order representation rank
exactly two on its complete original source.
For \(s=1\), all chains are binary quota-one; for \(s=2\), they
all have half quota.
On these query instances \(Z\) has a deterministic FPTAS, and
exact uniform sampling takes \(1+o(1)\) expected polynomial-time
proposals as the item count tends to infinity.
\end{theorem}

We isolate the only property needed by the list-pruning algorithm.
The hypotheses concern original-label means, not just defect totals.

\begin{lemma}[Deleting a color]\label{comp:deleting-color}\label{lem:deletion-front}
Let \(a,c\ge1\) and \(K\ge2\) be integers. Consider a complete
unit-weight source of \(g=4aK\) groups of the type in
Appendix~\ref{comp:gadget-section}, equally divided among \(K\)
colors, with defect polynomial
\[
 P_c(z)=\sum_{d=0}^c\binom cd^s z^d,\qquad s\in\{1,2\}.
\]
Assume every feasible word has at most \(a\) defects; every
feasible word with exactly \(a\) defects is monochromatic and
its color is soluble. Define \(\mu_r\) by assigning each group
of color \(r\) mean defect \(1/4\), uniformly over its original
labels, and making all other groups all high. Assume \(\mu_r\)
is a feasible original-label mean whenever color \(r\) is soluble.

Let \(zP_c'(z)/P_c(z)=1/(4K)\), \(L=-\log z\), and
\[
 \Delta=L-\frac gK\log\frac{P_c(ze^{L/a})}{P_c(z)}.
\]
If \(G(\beta)-\log U<\Delta\), deleting any color minimizing
\(\beta\cdot\mu_r\) preserves whether a soluble color exists.
For fixed \(K\) and \(c=a^{O(1)}\),
\[
 L=s\log c+O_K(1),\qquad
 \Delta=(1-1/K)L-O_K(L^2/a).
\]
\end{lemma}
\begin{proof}
Let \(m=1/(4K)\). By~\eqref{comp:odds-bounds},
\(m\le c^sz\le m/(1-m)\), so \(L>0\).
The product group law \(Q_z\) has original-label mean
\(K^{-1}\sum_r\mu_r\) and entropy
\[
 \mathcal S(a)=g\log P_c(z)+aL.
\]

\paragraph{If the sole soluble color minimizes the score, \(G(\beta)\ge\mathcal S(a)\).}
Only deletion of a unique soluble color \(r\) could change the
answer. If this color minimizes the score, its feasible mean
and the average-score inequality give
\[
 m_{\mathcal F}(\beta)\le\beta\cdot\mu_r
               \le\beta\cdot\mathbb E_{Q_z}\chi .
\]
The Gibbs inequality therefore yields
\begin{equation}\label{comp:wrong-delete}
 G(\beta)=\log P(\beta)-m_{\mathcal F}(\beta)\ge\mathcal S(a).
\end{equation}
This bound applies to every signed potential, without any
symmetry assumption on its coordinates.

\paragraph{The same instance has log capacity at most \(\mathcal S(a)-\Delta\).}
Assign potential \(-L\) per defect and reward each defect of
color \(r\) by \(u=L/a\). Words with at most \(a-1\) defects
have potential at least \(-(a-1)L\); a feasible word with \(a\)
defects must be monochromatic in color \(r\), and has exactly
that potential. Using \(-(a-1)L\) as a valid threshold and
evaluating on the complete product source gives
\begin{equation}\label{comp:unique-color}
 \log U
 \le \mathcal S(a)-L+
      \frac gK\{\log P_c(ze^u)-\log P_c(z)\}
 =\mathcal S(a)-\Delta .
\end{equation}
Equations~\eqref{comp:wrong-delete} and~\eqref{comp:unique-color}
contradict \(G(\beta)-\log U<\Delta\).
If there are at least two soluble colors, or none, any deletion
preserves the answer. This also covers ties.

\paragraph{Size of the deletion margin.}
Write \(p_d=\binom cd^s\) and \(V=c^s\). Iterating \(d\,p_d\le Vp_{d-1}\) gives
\[
 \mathbb E_{z'}d\le Vz',\qquad
 \mathbb E_{z'}[d(d-1)]\le(Vz')^2,
 \qquad \operatorname{Var}_{z'}(d)\le Vz'+(Vz')^2 .
\]
The odds bounds give \(L=s\log c+O_K(1)\).
For polynomially bounded \(c\), \(u=L/a\to0\), so the variance
is \(O_K(1)\) uniformly for \(z'\) between \(z\) and \(ze^u\).
The first and second derivatives of \(\log P_c(e^y)\) are
the mean and variance. Taylor's theorem in \(y=\log z\) has
linear term \((g/K)\,m\,u=L/K\) and remainder \(O_K(L^2/a)\),
which proves the claimed formula for \(\Delta\).
The real potential in~\eqref{comp:unique-color} is solely an
upper-bound witness; the decoder does not compute it.
\end{proof}

\begin{proof}[Proof of Theorem~\ref{comp:bare-hardness}]
\textbf{A query and its rational decoder.}
Choose fixed integers \(K\ge2\) and \(b\ge1\) such that
\[
 (1-1/K)\frac{sb}{b+1}>\gamma.
\]
This is possible for every \(\gamma<s\).
Encode \(K\) padded lists of common even length \(a\) by
Lemma~\ref{comp:gadget}. The bulk has at most \(a-1\) defects;
every corner word has \(a\) defects in one soluble color.
Lemma~\ref{comp:color-means} supplies the remaining hypothesis
of Lemma~\ref{comp:deleting-color}.

Given a promised rational output \(\beta\), let \(X_{\rm top}\)
be the all-high word. The scores have the explicit form
\begin{equation}\label{comp:rational-color-score}
 \beta\cdot\mu_r
 =\beta\cdot\chi(X_{\rm top})+
   \frac1{4c}\left(
      \sum_{i\in\mathrm{low}(r)}\beta_i-
      \sum_{i\in\mathrm{high}(r)}\beta_i\right).
\end{equation}
Delete a color minimizing the parenthesized rational sum, with
a fixed tie rule. This is exactly minimization of
\(\beta\cdot\mu_r\), whether or not the list is soluble. Since
\[
 \log N=(b+1)\log a+O_{K,b}(1),\qquad
 L=sb\log a+O_{K,b}(1),\qquad L^2/a=o(1),
\]
the choice of constants ensures \(\Delta>\gamma\log N\)
for all sufficiently large \(a\). The output guarantee implies
\(G(\beta)-\log U\le\gamma\log N\), so the deletion lemma
preserves the disjunction of the \(K\) answers.
No query to \(m_{\mathcal F}\), no evaluation of \(G(\beta)\)
or \(U\), and no computation of \(z,L,\Delta\) occurs in the decoder.

\paragraph{The list invariant and the number of queries.}
Maintain at most \(K-1\) unpadded lists of equal length whose
disjunction has the same answer as the original instance.
Initially there is just the input list. Replace the first two
entries \(x,y\) of each retained list by \(x+y\) in one child
and by \(|x-y|\) in the other. These are the equal-sign and
opposite-sign cases in the signed-sum formulation of
\textsc{Partition}, so the disjunction is preserved.

There are at most \(2(K-1)\) children. While at least \(K\)
are present, query any \(K\) and delete the color returned by
the rational rule. Each deletion preserves the disjunction of
that sublist, hence of all retained lists. At most \(K-1\)
queries restore the invariant. Append one zero only inside
a query if an even length is needed; the retained lists remain
unpadded. Below a fixed length cutoff, decide all remaining
lists by direct enumeration. Choose this cutoff large enough
that every issued query has \(a\ge4\) and
\(\Delta>\gamma\log N\).
The unpadded length decreases at every branching level.
There are linearly many levels and \(O(K)\) queries per level.

\paragraph{Bit complexity.}
Branching never increases a list's total weight.
For an input of bit length \(n\), \(a\le n+1\),
\(\log(1+W_*)=O(n)\), and \(c\) is polynomial in \(n\)
for fixed \(b\). The explicit query has polynomial bit length
by Lemma~\ref{comp:gadget}. Rational scores
\(\beta\cdot\mu_r\) can be compared in time polynomial in the
query and output lengths; a polynomial-time solver has
polynomial-length output. The cutoff depends only on the
fixed \(s,\gamma,K,b\), never on the numerical item weights.
This proves the deterministic polynomial-time Turing reduction.

\paragraph{Counting and sampling the same original-label family.}
The mass FPTAS follows from Theorem~\ref{comp:count-sample}:
each query has supplied rank two and unit activities.
For the stronger proposal bound, write
\[
 D_j=[x^j]P_c(x)^{g/2},\qquad
 B_0=\left(\sum_{j=0}^{a/2}D_j\right)^2-D_{a/2}^2.
\]
Here \(D_j\) counts the original-label choices on one sign
side, and \(B_0\) is the exact bulk size.
Truncated coefficient convolution computes these integers.
Draw the sign totals with probabilities proportional to \(D_iD_j\),
excluding \((a/2,a/2)\); conditional coefficient unranking
then draws the group defects. For a group with \(d\) defects,
choose its defective slots when \(s=1\), or independently choose
its \(d\) selected lows and \(d\) omitted highs when \(s=2\).
This samples the bulk uniformly, including the original labels.

The disjoint set of monochromatic one-defect-per-item candidates
has size \(C_0=K(4V)^a\). Draw uniformly by choosing a color
and, independently at each item, a side, a sign, and a one-defect
labelled pattern. The feasible family consists of the bulk and
a subset of these candidates. Mix the two sets in exact
proportion \(B_0:C_0\), and reject precisely when the original
rows fail. The accepted law is exactly uniform on \(\mathcal F\).

To bound the overhead, fix \(a-1\) items and put one defect on
each. Choose \(a/2\) of them to have positive sign; choose any
color, side, and local pattern independently for each item.
These distinct bulk words give
\[
 B_0\ge\binom{a-1}{a/2}(2KV)^{a-1}
       \ge\frac{2^{a-1}}a(2KV)^{a-1},\qquad
 \frac{C_0}{B_0}\le\frac{4KaV}{K^{a-1}}=o(1).
\]
The expected number of proposals is at most
\((B_0+C_0)/B_0=1+o(1)\).
Coefficient tables and unranking have polynomial size, with
entries of at most \(N+1\) bits because they count source words.
Exact rational choices use uniform integers and fair-bit
rejection, giving expected polynomial bit time.
\end{proof}

The reduction extracts a decision from the potential's coordinates,
not from an evaluated envelope. The sharp exponent here uses
adaptive list pruning; the one-query value reductions in
Appendix~\ref{comp:value-section} have a separate decoder.

%% file: proofs/value_hardness.tex
\section{Sharp value exponents at intrinsic rank two}
\label{comp:value-section}

\begin{theorem}\label{comp:value-hardness}
For each \(s\in\{1,2\}\), the following tasks are NP-hard on the
two-row, unit-activity, minimum-rank-two family of
Lemma~\ref{comp:gadget}:
\begin{enumerate}
\item returning a rational two-sided \(N^\gamma\) approximation
of the all-cut capacity, for every fixed \(0<\gamma<s/2\);
\item returning a certified upper value within \(N^\gamma\) of
the capacity, for every fixed \(0<\gamma<s\).
\end{enumerate}
For \(s=1\) all chains are binary quota-one; for \(s=2\) all
have half quota. Both score rows are strictly increasing on
each chain. Each reduction uses one query. The same instances
have the true-count FPTAS and exact sampler of
Theorem~\ref{comp:count-sample}.
\end{theorem}
\begin{proof}
Reduce from \textsc{Partition}~\cite{Karp72}. Pad its list with
zeros to an even length \(a\ge4\), and use
Lemma~\ref{comp:gadget} with \(K=1\) and a fixed exponent
\(b\) to be specified. Put \(g=4a\), and define
\[
 \mathcal S(j)=gH_c(j/g)
 \qquad(0<j<gc).
\]
In a NO instance the maximal feasible defect total is \(a-1\);
in a YES instance it is \(a\).

The uniform law on the first bulk layer
\(\sum d=a-1\), \(|S|=1\), has every group mean
\((a-1)/g\). The bulk ignores items and side choices,
so permutations within each sign and exchange of the signs
give this equality. Its individual-label means equal those
of the product group Gibbs law with that mean.
The entropy certificate therefore gives
\(\log U\ge\mathcal S(a-1)\) in both branches.
In a YES instance, fix one partition solution, mix it fairly
with its complement, independently choose a uniform balanced
sign vector, and choose each one-defect labelled pattern
uniformly. This is a feasible corner law with mean defect
\(1/4=a/g\) in every group and with the same individual-label
means as the corresponding product Gibbs law. It gives
\(\log U\ge\mathcal S(a)\).

These lower bounds are exact. At total mean \(j\), let \(z_j\)
solve \(m(z_j)=j/g\). Inequality \eqref{comp:odds-bounds}
shows \(z_j<1\) for \(j\in[a-1,a]\).
Assign potential \(\log z_j\) to each selected low label and
zero to each high label. A word has potential
\((\sum d)\log z_j\), whose minimum over a branch with
maximal defect \(j\) is \(j\log z_j\).
The resulting log envelope is \(\mathcal S(j)\).
Consequently
\begin{equation}\label{comp:branch-values}
 \log U=
 \begin{cases}
 \mathcal S(a),&\textsc{Partition} \text{ is YES},\\
 \mathcal S(a-1),&\textsc{Partition} \text{ is NO}.
 \end{cases}
\end{equation}

For \(j\in[a-1,a]\), the group mean belongs to
\([3/16,1/4]\). Thus \eqref{comp:odds-bounds} gives
\(\mathcal S'(j)=-\log z_j=s\log c+O(1)\),
with an absolute uniform error. The branch log gap \(\Delta\)
therefore satisfies
\[
 \Delta=\mathcal S(a)-\mathcal S(a-1)
          =s\log c+O(1),\qquad
 \frac{\Delta}{\log N}\longrightarrow\frac{sb}{b+1}.
\]

For the comparison we use finite rational arithmetic, without treating the
NO value as an exact real constant. For \(m_0=(a-1)/g\), its
odds lie in \([1/(8V),1/V]\) by
\eqref{comp:odds-bounds}. Bisection using the rational sign of
\(zP_c'(z)-m_0P_c(z)\) locates the root in a rational interval
of relative width at most \(1/(4gc)\).
At either endpoint evaluate exactly the positive rational
\[
 R_0=P_c(z)^g z^{-(a-1)}.
\]
The function \(\log P_c(z)^g-(a-1)\log z\) is minimized at the
root and has derivative in \(\log z\) of magnitude at most
\(gc\). Hence
\begin{equation}\label{comp:no-reference}
 U_{\rm NO}\le R_0\le e\,U_{\rm NO}.
\end{equation}
All coefficients, evaluations, and integer powers have polynomial
bit length: \(c\) and \(g\) are polynomial, and only
\(O(\log(Vgc))\) bisections are needed for this relative width.

For a two-sided exponent \(\gamma<s/2\), choose fixed \(b\)
large enough that \(sb/(b+1)>2\gamma\).
Choose a fixed rational \(\delta>\gamma\) with
\(\delta+\gamma<sb/(b+1)\).
Given a promised value \(Y\), compare it to \(R_0N^\delta\).
On NO instances \(Y\le U_{\rm NO}N^\gamma<R_0N^\delta\).
On sufficiently large YES instances,
\[
 Y\ge U_{\rm YES}N^{-\gamma}
 \ge R_0\exp(\Delta-1)N^{-\gamma}>R_0N^\delta.
\]
The comparison is exact rational arithmetic after raising both
sides to the denominator of \(\delta\).
Bounded \(a\) is handled by direct enumeration.

For a certified upper value with exponent \(\gamma<s\),
choose \(b\) and a rational \(\delta>\gamma\) with
\(\delta<sb/(b+1)\), and use the same threshold.
NO outputs are at most \(U_{\rm NO}N^\gamma\).
YES outputs are at least \(U_{\rm YES}\), which exceeds the
threshold for large \(a\).
This uses only the guaranteed upper value, so it also excludes
the stronger output of a valid rational potential and threshold
with such a value. These one-query reductions preserve the
minimum-rank promise in both branches by
Lemma~\ref{comp:gadget}.
\end{proof}

%% file: proofs/forest.tex
\section{Capacity, mass, and exact sampling on incidence forests}\label{app:forest}
We prove Theorem~\ref{thm:forest-main}. If the source has only one word, test it directly. If it is feasible, its rational weight gives the mass and the certified value at the zero field, and the sampler returns that word. In the remaining cases, remove frozen chains, subtract their exact row contributions, and retain their activity product. Test isolated rows as constants. In a scalar chain, absorb $g_cb_{jc}$ into the threshold and write its row contribution as $\alpha_{jc}S_c(X_c)$, where $S_c(X_c)=\sum_{i\in X_c}s_i$ and $\alpha_{jc}\ge0$. To recognize a scalar chain, use any nonconstant incident profile as the coordinate. Check by exact rational arithmetic that every other profile is affine in that coordinate with nonnegative slope. A chain with no incident row uses a constant coordinate. A complete enumerated domain stores one state for each original quota word, with its activity and exact row contributions.

\subsection{An exactly valid additive support oracle}
Let $[0,R]^N$ be the finite numerical box in~\eqref{comp:box},
with $N$ still counting all original labels. Let $I$ denote the
nonfrozen labels, and let $c_f(\beta)$ be the exact potential of
the forced selected labels, as in Appendix~\ref{comp:finite}.
The recurrences below process only $I$ and restore the frozen
selections in the final witness.
Given a rational $\beta$ in this box and a tolerance $\delta>0$, put $\eta=\delta/N$ and assign each $i\in I$ the nonnegative integer cost $k_i=\lfloor\beta_i/\eta\rfloor$. Every full word has
\begin{equation}\label{eq:forest-cost}
 c_f(\beta)+\eta\sum_{i\in X\cap I}k_i\le\beta\cdot\chi(X)
 <c_f(\beta)+\eta\sum_{i\in X\cap I}k_i+\delta.
\end{equation}
The total active integer cost is at most $N^2R/\delta$.

For a scalar chain process labels with a take/skip dynamic program indexed by prefix, selected count, and integer cost. At each state retain the largest attainable exact scalar sum and an original-label predecessor. On the quota layer this gives at most one representative per integer cost. Replacing a local word by this representative preserves its rounded cost and improves all incident row contributions simultaneously. Thus replacing words chain by chain preserves feasibility. Conversely every representative is an actual local word. This compression preserves the exact minimum rounded cost on the global feasible family; it does not replace the source partition $P$ in the capacity objective. For an enumerated chain keep every local state and its rounded cost.

After checking isolated rows, root every component at a chain, including each isolated chain. At a chain $c$ and a fixed local state $u$, its message to the parent row is its local cost plus the child-row minimum costs for this same state. A missing feasible child choice gives infinity. At a row $j$, each child chain supplies pairs consisting of a subtree cost and an exact contribution to row $j$. Convolve these lists by total integer cost, retaining the largest total row contribution at each cost. For each parent-chain state, subtract its contribution from the row threshold and take the least cost reaching that residual threshold. Keep backpointers. A row with no children has one aggregate of cost and score zero. If a child has no feasible state, the corresponding branch is infeasible.

Deleting a parent edge separates the child subtrees. Therefore, by induction, the chain and row recurrences compute their exact minima on the compressed local domains. No label cost is counted twice, and each incident row uses the same state at a chain. Minimize over the root states and add component minima. If $J$ is the resulting rounded minimum, backtracking gives $X\in\F$ and
\begin{equation}\label{eq:forest-oracle}
 t=c_f(\beta)+\eta J\le m_\F(\beta)\le\beta\cdot\chi(X)\le t+\delta.
\end{equation}
All comparisons with row thresholds use exact rational scores. The cost ranges, enumerated domain sizes, and scalar tables are polynomial in $\bits,D,1/\delta$, the binary encoding lengths of the rational query $(\beta,\delta)$, and the numerical box radius. Exact sums of input rationals and the scalar coordinates have polynomial bit length. Thus~\eqref{eq:forest-oracle} is an additive support oracle with an exactly valid lower threshold. The finite-box optimizer of Lemma~\ref{comp:finite-optimizer}, followed by its last support call and an outward rational evaluation, gives the capacity FPTAS.

\subsection{A precise one-dimensional trimming operation}
For a finite nonnegative atomic measure $\mu$ on rational scores, set $H(x)=\mu([x,\infty))$. Return the zero measure unchanged. Otherwise list its positive-mass scores in decreasing order, and let $h_j$ be their cumulative masses. Retain the first positive cumulative value, and subsequently retain $h_j$ exactly when it exceeds $(1+\eta)$ times the last retained value. At an original score location that is not retained, use the preceding retained cumulative value. Define values between original scores by the corresponding weak upper tail, and use zero above the largest score. The differences of the retained cumulative values are nonnegative rational masses at the retained \emph{original score locations}. This defines an atomic measure $\widetilde\mu$ and handles ties by first aggregating equal scores.

\begin{lemma}[Tail comparison]\label{lem:forest-trim}
For every real $x$,
\[
 H(x)/(1+\eta)\le\widetilde\mu([x,\infty))\le H(x).
\]
The same multiplicative comparison holds when integrating any nonnegative increasing function of the score. Arbitrary shifts, nonnegative scalar rescaling, addition, and convolution with a fixed nonnegative measure preserve the corresponding comparison.
\end{lemma}
\begin{proof}
The zero case is immediate. At a retained score the cumulative value is exact. At every discarded score the current cumulative value is at most $(1+\eta)$ times the preceding retained value, by the retention rule. Between scores the same tails remain constant. A nonnegative increasing function on the union of the two finite supports is a nonnegative linear combination of a constant and indicators of weak upper intervals; summation proves the integration assertion. Shifts and scaling preserve these intervals. Addition preserves inequalities, and convolution integrates a tail at shifted thresholds against a nonnegative measure. Commutativity gives the corresponding assertion for either argument of a convolution. Applying it successively to both arguments multiplies their error factors and preserves exact zero tails.
\end{proof}

This is a standard one-dimensional approximation-set operation, closely related to knapsack counting schemes~\cite{Halman2016Forest}. We need its one-sided form and an accounting of its use on the original quota source.

\subsection{Scalar distributions and tree messages}
For each scalar chain maintain one atomic measure per cardinality while processing its labels. Start with unit mass at score zero and cardinality zero. A take transition shifts the preceding-cardinality measure by $s_i$ and multiplies it by $\lambda_i$; a skip transition preserves the current-cardinality measure. Add these two measures, then trim each resulting layer once. Induction using Lemma~\ref{lem:forest-trim} shows that after $k$ labels every layer's tails approximate their exact values within $(1+\eta)^k$. Addition of alternatives takes the larger error exponent, rather than their sum. Keep the final quota measure.

Hold all other chain states fixed. Feasibility as a function of a scalar chain's score is a weak upper interval, possibly empty or the whole line, because every incident coefficient $\alpha_{jc}$ is nonnegative. Replacing the scalar-chain measures one at a time therefore decreases the total feasible mass by at most a factor $(1+\eta)^N$. This replacement argument remains valid after earlier chains have already been replaced by arbitrary positive measures. Enumerated chains retain their exact local masses and vector scores. Write $\overline Z$ for the exact feasible mass on these finite local domains, including the retained frozen activity product. Then
\begin{equation}\label{eq:forest-local-error}
 \overline Z\le Z\le(1+\eta)^N\overline Z.
\end{equation}
The trimmed jumps serve only as counting weights, not as probabilities assigned to recovered quota words.

On these finite domains perform tree sum-product~\cite{KschischangFreyLoeliger2001}. Conditional on a chain state, multiply its local mass by the child-row messages evaluated at that same state. At a row, push each child-chain measure forward to its contribution to that row, start with unit mass at zero, convolve the children sequentially, and trim after each convolution. The resulting survival function, evaluated at the threshold minus the parent contribution, serves every parent state. At each root sum state weights; multiply results over components and by the retained frozen activity product.

Here is the pointwise invariant for the second approximation stage. Let $M_{v\to u}$ denote an exact message on the finite local domains, and $\widehat M_{v\to u}$ its computed message. A chain-to-row message is indexed by that chain's local state and includes its local mass; a row-to-chain message is indexed by the parent state and excludes the parent chain's mass. For every state at the boundary,
\begin{equation}\label{eq:forest-message-invariant}
 (1+\eta)^{-e_{v\to u}}M_{v\to u}
 \le\widehat M_{v\to u}\le M_{v\to u},
\end{equation}
where $e_{c\to j}=\sum_{h\text{ child row of }c}e_{h\to c}$ and
$e_{j\to c}=\sum_{d\text{ child chain of }j}e_{d\to j}+k_j$, with $k_j$ the number of child chains of row $j$.
At a leaf chain the message is its exact local mass. A childless row returns the exact zero-or-one threshold test. At a chain, multiply the child bounds at the \emph{same} local state. At a row, the child bounds give pointwise comparisons of the pushed-forward atomic weights, and hence the same tail comparisons. Sequential positive convolutions preserve the tail comparisons; each of the $k_j$ trims introduces one factor by Lemma~\ref{lem:forest-trim}. After a trim the comparison concerns tails, not individual atomic weights. Tail evaluation gives the pointwise bound simultaneously at every parent state, including exact threshold ties. This proves~\eqref{eq:forest-message-invariant} by induction without any independence assumption on different parent queries.

Root summation preserves each component's factor, and products over components add the exponents. Each row--child-chain edge is charged once, so their total is at most the incidence count $E$. Combining this bound with~\eqref{eq:forest-local-error}, and writing $L$ for the algorithm's answer, gives
\[
 L\le Z\le(1+\eta)^{N+E}L.
\]
Choose $\eta=\varepsilon/[2(1+N+E)]$ for $0<\varepsilon\le1$. Since $e^{\varepsilon/2}\le1+\varepsilon$, this is the desired one-sided FPTAS. Infeasible residuals have exact zero messages.

\paragraph{Bit complexity.}
For a set of labels $S$, let $D_S$ be the product of the denominators of their reduced rational activities. Every mass on a prefix or subtree is a nonnegative integer multiple of $1/D_S$. Trimming copies cumulative values and takes differences. Multiplication at a chain or row uses disjoint label sets; a conditional row message excludes the parent chain's activity. The total mass is bounded by $\prod_{i\in S}(1+\lambda_i)$. Hence positive cumulative values have polynomial logarithmic range, and a trimmed list has
\[
 O\!\left(1+\frac{\log(D_S\prod_{i\in S}(1+\lambda_i))}{\log(1+\eta)}\right)
\]
entries. Scores are sums of input-derived rationals and have polynomial bit length. A convolution enumerates pairs from two polynomial lists, aggregates exact equal scores, sorts, and trims. These bounds apply at arbitrary row degree. There is no loop indexed by a binary-encoded score magnitude.

\subsection{Restoring the exact original-label law}
Process the original labels in any fixed order. A prefix fixes some bits, updates the remaining quotas and row thresholds, and contributes its forced activity product. Scalar restrictions remain scalar; enumerated domains can be filtered; removing decided labels cannot introduce a cycle. Thus every valid residual quota instance remains in the same class. An invalid quota gives zero mass. An empty completion has unit residual mass when it meets the residual thresholds, and this mass is multiplied by the forced activity factor.

Let $Z_b$ be the weighted completion mass of child bit $b$ at a prefix, including the forced factor. Compute lower estimates $L_b$ with $Z_b/(1+\rho)\le L_b\le Z_b$, where $\rho=1/(4N)$. Zeros are recognized exactly. Choose the next bit with rational probability $L_b/(L_0+L_1)$. Whenever that branch has positive mass,
\[
 \frac{L_b}{L_0+L_1}\ge\frac{1}{1+\rho}\frac{Z_b}{Z_0+Z_1}.
\]
The resulting computable proposal probability $Q(X)$ therefore satisfies
\[
 Q(X)\ge(1+\rho)^{-N}\frac{w(X)}Z\ge\frac{w(X)}{2Z}
 \qquad(X\in\F).
\]
One root FPTAS call at tolerance one returns $L_{\rm root}$ with $Z/2\le L_{\rm root}\le Z$; set $B=2L_{\rm root}$. Accept a proposed word with the exactly computable rational probability $w(X)/(2BQ(X))$. This belongs to $[0,1]$ and assigns unconditional probability exactly $w(X)/(2B)$ to returning $X$ in one trial. Thus accepted words have the exact target law and the expected number of trials is $2B/Z\le4$. Each rejected trial is restarted with fresh fair bits. Rational choices are implemented with uniform integer generation by fair-bit rejection. The path probabilities and their products have polynomial bit length, and each trial makes only polynomially many FPTAS calls at tolerance $1/(4N)$. This proves expected polynomial bit time with no output approximation.

\subsection{The binary width threshold}
Every binary quota-one chain has an explicit two-state domain. Thus the positive theorem applies to every supplied binary incidence forest. The two-row binary queries in the bare-potential reduction have both rows nonconstant on every chain, so their graph is $K_{2,d}$ with $d\ge2$. Bags containing both rows and one chain form a width-two decomposition. A four-cycle excludes width one. Theorem~\ref{thm:separation-main}(ii) therefore proves the claimed width-two capacity barrier. It is a hardness result for this width-two class, and implies no counting lower bound.

%% file: proofs/single_chain.tex
\section{A single long chain: an exact shell certificate}\label{app:single-chain}
We prove the first assertion of Theorem~\ref{thm:path-main}, with $c=1/100$. The central construction has two certificates: a complementary feasible pair on YES, and a feasible realization of the optimal grade tilt's original-label means on NO. These give the capacity gap and its rational decoder. The stencil's independent rank-two certificate is checked afterwards.

\subsection{The two increasing rows}
Start with a nonnegative integer instance $w_1,\ldots,w_a$ of \textsc{Partition}~\cite{Karp72}, of total $W$. Zero padding makes $a$ even and at least two. Place one dummy quadruple, then the $a$ item quadruples, then a second dummy quadruple, in a single chain. Write
\[
 n=4(a+2)\ge16,\quad d=n/2,\quad
 \Omega=\binom{[n]}d,\quad B=\binom nd,\quad
 G_0(X)=\sum_{i\in X}i,\quad \mu=d(n+1)/2.
\]
Even padding makes $n$ a multiple of eight and $d$ a multiple of four. All activities are one. Put $h_i=w_j$ at the first position of item quadruple $j$, and zero at other positions. Let the stencil $s$ be $(1,-1,-1,1)$ on the first dummy quadruple, its negative on the last, and zero elsewhere. Put
\[
 H(X)=\sum_{i\in X}h_i,\quad S(X)=\sum_{i\in X}s_i,\quad
 J=2W+1,\quad C=3J.
\]
The two label scores and thresholds are
\begin{equation}\label{eq:stencil-rows}
 a_\pm(i)=Ci\pm(2h_i+Js_i),\qquad
 \theta_\pm=C\mu\pm W.
\end{equation}
Every consecutive score difference is at least $C-2W-2J=1$, so both rows strictly increase. Coefficients and thresholds have $O(\log(n+1)+\log(W+2))$ bits. The complete source is the single half slice; we impose no per-quadruple quota.

Feasibility is exactly
\begin{equation}\label{eq:stencil-test}
 C(G_0(X)-\mu)\ge |2H(X)-W+JS(X)|.
\end{equation}
Since $|S|\le4$ and $|2H-W|\le W<J$, every grade below $\mu$ fails, every grade at least $\mu+2$ passes, and a grade-$\mu+1$ word with $|S|\le2$ passes. At grade $\mu$, integrality forces $S=0$ and $2H=W$. Thus any feasible central word yields a partition from its selected item-first positions, including words with irregular quadruple occupancies.

Conversely, denote the outer and inner pairs of a quadruple by $O=\{1,4\}$ and $I=\{2,3\}$ in local positions. Both have the same grade and cardinality. For a partition, choose $O$ on its selected items and $I$ elsewhere, with $O$ at both dummy ends. This word and its complement satisfy~\eqref{eq:stencil-test} at grade $\mu$. Their mean is $p_0=\mathbf1/2$, the uniform source mean. Proposition~\ref{prop:entropy} and the zero potential give
\begin{equation}\label{eq:single-yes}
 U=B\qquad\text{on YES}.
\end{equation}

\subsection{A flow certificate on a guaranteed feasible shell}
The capacity lower bound needs a feasible law with prescribed
\emph{original-label} means. We construct it inside the smaller family
\[
 \mathcal T=\{X\in\Omega:G_0(X)=\mu+1,\ |S(X)|\le2\}\subseteq\F.
\]
This inclusion holds for every item list, since
$|2H-W+JS|\le W+2J<C$ on $\mathcal T$. Thus the construction
below need only reach grade $\mu+1$ and control $S$; it need not
preserve the encoded statistic $H$.

Let $e_i$ be the $i$th coordinate vector. Put $p_0=\mathbf1/2$,
$b_i=e_{i+1}-e_i$ for $1\le i<n$, and
$U_i(X)=(X\setminus\{i\})\cup\{i+1\}$ when $i\in X$ and $i+1\notin X$. It changes the incidence vector by $b_i$, the
grade by one, and the stencil sum by at most two. Also put
$R(X)=\{n+1-i:i\in X\}$ and $\tau(X)=[n]\setminus R(X)$.
The half quota and the antisymmetric stencil give
\begin{equation}\label{eq:shell-reflection}
 G_0(\tau(X))=G_0(X),\qquad S(\tau(X))=S(X).
\end{equation}
In particular $\tau(\mathcal T)=\mathcal T$. No invariance of
$H$ or of the whole feasible family is asserted.

\begin{lemma}[Surviving-shell means]\label{lem:shell-flow}
Let $t_1,\ldots,t_{n-1}$ be nonnegative, with
\[
 \sum_i t_i=1,\qquad t_i=t_{n-i},\qquad t_{d-1}\ge t_d/2.
\]
Then $p_0+\sum_i t_i b_i$ is the mean of a law on at most $n+2$
actual feasible words of grade $\mu+1$ in~\eqref{eq:stencil-test},
for every encoded item list. Its weights are explicit linear
functions of $t$. For rational $t$, the law can be constructed in
time polynomial in $n$ and the encoding length of $t$.
\end{lemma}
\begin{proof}
We first construct laws for the paired bonds $i,n-i$, then bridge
the central bond $d$, and finally mix them.

\paragraph{Paired bonds.}
For $1\le i<d$, let $X_i$ be a union of $d/2$ reflection pairs,
containing $\{i,n+1-i\}$ and omitting $\{i+1,n-i\}$.
The remaining $d/2-1$ pairs can be the first available ones:
there are $d-2\ge d/2-1$ choices, even for $i=d-1$.
Both $X_i$ and its complement have grade $\mu$ and stencil sum
zero. The fair law $L_i$ on
\[
 U_i(X_i),\qquad U_{n-i}([n]\setminus X_i)
\]
is supported on $\mathcal T$ and has mean
$p_0+(b_i+b_{n-i})/2$.

\paragraph{The central bridge.}
A union of reflection pairs occupies $d$ and $d+1$ together, so
it cannot support the upgrade $U_d$. Since $d$ is divisible by four, take the outer pair
in every quadruple except the first quadruple right of $d$, where
we take the inner pair. This exceptional quadruple is an item
quadruple for $n\ge16$, and both dummy ends retain outer pairs.
The resulting $X_*$ has grade $\mu$, stencil sum zero, and bits
$0,1,0$ at $d-1,d,d+1$. Hence the fair law $L_-$ on
\[
 U_d(X_*),\qquad U_{d-1}([n]\setminus X_*)
\]
is supported on $\mathcal T$: the changed positions lie in item
quadruples, so $S$ remains zero. Its mean is
$p_0+(b_{d-1}+b_d)/2$. Let $L_+$ be its image under $\tau$.
Equation~\eqref{eq:shell-reflection} keeps both atoms in
$\mathcal T$, and $\tau$ sends a displacement $x$ from $p_0$
to $-Rx$. Since $-Rb_i=b_{n-i}$, the mean of $L_+$ is
$p_0+(b_d+b_{d+1})/2$.

\paragraph{Mixture and labelwise verification.}
Use the two central laws with weight $t_d$ each. They supply all
of $t_d b_d$ but also $t_d/2$ at each neighboring bond. Subtract
those contributions from the ordinary law $L_{d-1}$:
\begin{equation}\label{eq:shell-mixture}
 \nu_t=\sum_{i=1}^{d-1}
       \bigl(2t_i-t_d\ind_{\{i=d-1\}}\bigr)L_i
       +t_dL_-+t_dL_+.
\end{equation}
The neighbor condition is exactly what makes the adjusted
coefficient nonnegative. Symmetry gives total weight
$2\sum_{i<d}t_i+t_d=1$. The preceding means now give
$\E_{\nu_t}\chi=p_0+\sum_i t_i b_i$; equivalently, with
$t_0=t_n=0$,
\begin{equation}\label{eq:shell-coordinate}
 \mathbb E_{\nu_t}\chi_j=\tfrac12+t_{j-1}-t_j
 \quad(1\le j\le n).
\end{equation}
This verifies every label, not only the grade. The mixture has
$2(d-1)+4=n+2$ displayed atoms before coalescing repetitions.
Constructing their subsets and the linear rational weights has
the claimed polynomial bit cost.
\end{proof}

\paragraph{The central bridge at the smallest size.}
For $n=16$ and $(w_1,w_2)=(1,2)$, we have $d=8$, $\mu=68$,
$W=3$, $J=7$, and $C=21$. The construction uses
$X_*=\{1,4,5,8,10,11,13,16\}$. The following four atoms all
have grade $69$ and stencil sum zero:
\begin{center}
\begin{tabular}{@{}lll@{}}
\toprule
Law & Original-label word & $H$\\
\midrule
$L_-$ & $\{1,4,5,9,10,11,13,16\}$ & $3$\\
$L_-$ & $\{2,3,6,8,9,12,14,15\}$ & $2$\\
$L_+$ & $\{2,3,5,9,10,11,14,15\}$ & $3$\\
$L_+$ & $\{1,4,6,7,10,12,13,16\}$ & $0$\\
\bottomrule
\end{tabular}
\end{center}
Each law gives its two atoms probability $1/2$. Their means are
$p_0+(e_9-e_7)/2$ and $p_0+(e_{10}-e_8)/2$, respectively.
Complement-reflection maps the second atom of $L_-$ to the second atom of $L_+$, changing $H$ from $2$ to $0$ in this
pair, but all four atoms satisfy $|2H-3|\le3<C$.
The preserved quantities $G_0$ and $S$ therefore suffice for
actual feasibility.

\subsection{The exact capacity of the NO family}
Under the uniform source let $Y=G_0-\mu$, $K(u)=\log\E e^{uY}$, and let $Q_u$ be the full-source law proportional to $e^{uY}$. The distribution of $Y$ is symmetric by complementation, and its largest value is $n^2/8>1$. For every finite $u$, $K''(u)=\Var_{Q_u}(Y)>0$, since the tilted source has full support. Moreover $K'(0)=0$ and $K'(u)\to n^2/8$ as $u\to\infty$. Continuity and strict monotonicity therefore give a unique finite $v>0$ with $K'(v)=1$. Put $p_i=Q_v(i\in X)$. To use Lemma~\ref{lem:shell-flow}, it remains to express $p-p_0$ as a symmetric nonnegative unit flow satisfying its central neighbor condition.

For $i<j$, swapping those labels shows $p_i<p_j$: the probabilities of the two exclusive patterns are in ratio $e^{v(j-i)}$ in favor of $j$. Complement-reflection preserves the tilted law, so $p_i+p_{n+1-i}=1$. Define
\[
 t_i=\sum_{j=1}^i(1/2-p_j),\qquad 1\le i<n.
\]
With $t_0=t_n=0$, telescoping gives
$p_j=1/2+t_{j-1}-t_j$. Reflection symmetry gives $t_i=t_{n-i}$.
For $i\le d$, the increments $1/2-p_i$ are positive and decreasing,
so the flows are nonnegative and
$t_d-t_{d-1}\le t_d/d$. Thus
$t_{d-1}\ge(1-1/d)t_d\ge t_d/2$. Summation by parts gives
\[
 \sum_{i=1}^{n-1}t_i
 =\sum_{j=1}^n j(p_j-1/2)
 =\mathbb E_{Q_v}G_0-\mu=K'(v)=1.
\]
Equations~\eqref{eq:shell-mixture}--\eqref{eq:shell-coordinate}
therefore realize the exact original-label marginals of $Q_v$
on the surviving feasible first shell. The real number $v$ is
used only to prove this identity; the reduction does not compute it.

On NO, no central word is feasible and the minimum feasible grade is $\mu+1$, by the same shell lemma. The grade field $\beta_i=vi$ has tight cost
\[
 \log B+K(v)-v.
\]
The matching feasible mean and Proposition~\ref{prop:entropy} give the reverse bound, because $H(Q_v)=\log B+K(v)-vK'(v)$. Hence
\begin{equation}\label{eq:single-exact}
 U=B\exp(K(v)-v),\qquad K'(v)=1,
 \qquad\text{on NO}.
\end{equation}
The same value is the capacity of the full grade tail $\{G_0\ge\mu+1\}$, although the actual stencil family deletes some first-shell words. The equality is justified by the surviving marginal certificate.

\paragraph{Asymptotic size and a finite gap.}
On the uniform half slice the diagonal indicator variance is $1/4$, and off-diagonal covariance is $-1/[4(n-1)]$. On the zero-sum subspace this is $c_n I$, where $c_n=n/[4(n-1)]$. With $g_i=i-(n+1)/2$ and $L=\sum_i g_i^2=n(n^2-1)/12$, it follows that
\begin{equation}\label{eq:single-variance}
 \sigma^2=\Var(Y)=c_nL=\frac{n^2(n+1)}{48}.
\end{equation}
Uniformly for $0\le u\le2/\sigma^2$, $|uY|=O(n^{-1})$. Symmetry and Taylor bounds give
\[
 K'(u)=\frac{\E[Y\sinh(uY)]}{\E\cosh(uY)}
 =u\sigma^2(1+O(n^{-2})).
\]
Its value at $2/\sigma^2$ exceeds one for sufficiently large $n$, so $v=\sigma^{-2}(1+O(n^{-2}))$. Integrating the same uniform estimate gives
\begin{equation}\label{eq:single-asymptotic}
 \log(B/U)=v-K(v)=\frac{1+O(n^{-2})}{2\sigma^2}
 =\frac{24+o(1)}{n^3}.
\end{equation}
The exact denominator in~\eqref{eq:single-variance} is retained in the relative-error expression.

We also need a finite gap for the decoder. At $u=1/\sigma^2$, $u\max|Y|=6/(n+1)<1/2$. For $|z|\le1/2$, Taylor's formula gives $\cosh z\le1+z^2\cosh(1/2)/2\le1+5z^2/8$. Consequently $K(u)\le5/(8\sigma^2)$ and
\begin{equation}\label{eq:single-finite-gap}
 \log(B/U)\ge u-K(u)\ge\frac{3}{8\sigma^2}
 =\frac{18}{n^2(n+1)}>\frac{16}{n^3}\qquad(n\ge16).
\end{equation}

\subsection{Decoding all three output contracts}
Let $\kappa=1/(100n^3)$ and suppose a rational field $\beta$ has tight log regret at most $\kappa$. Put $\beta^\circ=\beta-\bar\beta\mathbf1$, $r=\|\beta^\circ\|_2$, and
\[
 J_\beta=\log\E\exp\bigl(\beta\cdot(\chi-p_0)\bigr).
\]
Complement symmetry and $\cosh z\ge1+z^2/2$ give
\begin{equation}\label{eq:single-norm-lower}
 J_\beta\ge\log(1+c_nr^2/2)\ge0.
\end{equation}
On YES the feasible midpoint $p_0$ yields $G(\beta)\ge\log B+J_\beta$. Therefore $r^2\le2(e^\kappa-1)/c_n\le16\kappa<n^{-3}$.

On NO use the parabolic flow $t_i=i(n-i)/(2L)$. It has total one, is symmetric, and satisfies the neighbor condition in Lemma~\ref{lem:shell-flow}; its displacement is $\delta=g/L$. Thus $p_0+\delta\in\conv\chi(\F)$ and $\|\delta\|_2=1/\sqrt L<4n^{-3/2}$. It follows that
\[
 \log B-G(\beta)\le\beta^\circ\cdot\delta-J_\beta
 \le r/\sqrt L.
\]
By~\eqref{eq:single-finite-gap} a field of regret at most $\kappa$ saves more than $15/n^3$. Hence $r^2>n^{-3}$. The exact rational comparison of $n^3\|\beta-\bar\beta\mathbf1\|_2^2$ to one decodes the instance. Neither a support query nor an exponential evaluation is used.

A multiplicative factor $1+\kappa$ implies the required log regret, and a certified upper output supplies such a bare field. For a symmetric value $V$, compare with the rational threshold $B(1-n^{-3})$. On YES, $V\ge B/(1+\kappa)>B(1-n^{-3})$. On NO,~\eqref{eq:single-finite-gap} gives $V\le(1+\kappa)Be^{-16/n^3}<B(1-n^{-3})$; for example use $e^{-z}\le1-z/2$ for $0\le z\le1$. All comparison integers and rationals have polynomial bit length. This proves one-query NP-hardness for each output contract.

\subsection{An unconditional rank-two obstruction}
In the ordinary item blocks, take $I$ everywhere except at one block, where we take the local pair $\{2,4\}$. This raises the grade by one and leaves $H=0$. At the two dummy ends compare the four choices $OO,II,OI,IO$. All have grade $\mu+1$. The first two have $S=0$ and pass; the last two have $S=4,-4$ and fail, since $4J-W>C$. Coordinate by coordinate,
\[
 \chi(OO)+\chi(II)=\chi(OI)+\chi(IO).
\]
A single additive threshold cannot accept both first words and reject both last words: summing the two accepted inequalities contradicts summing the two strict rejected inequalities.

In any rank-one common-order representation on the same source, all nonconstant score rows are proportional modulo chain constants. The proportionality coefficients have the same positive sign: an oriented nonzero exchange increasing one row cannot decrease another in a common order. The quotas absorb the constants, and the strongest scalar threshold subsumes all rows. Constant rows either accept every word or make the family empty. Therefore a nonempty rank-one represented family is one additive halfspace. The trade excludes it, even if a new common order is chosen. The two displayed rows prove $q_{\min}(\F)=2$ on either outcome.

Under the uniform law on the complete half slice, each row in~\eqref{eq:stencil-rows} has mean $\theta_\pm$, since $\E G_0=\mu$, $\E H=W/2$, and $\E S=0$.
For $t_+,t_-\ge0$, set $\beta=t_+a_++t_-a_-$ and $\theta=t_+\theta_++t_-\theta_-$.
Then $\E[\beta\cdot\chi(X)]=\theta$, so Jensen's inequality gives the fixed-threshold Chernoff bound $P(\beta)e^{-\theta}\ge B$, with equality at zero multipliers.
On NO, the choice $t_+=t_-=v/(2C)$ gives $\beta_i=vi$ and $\theta=v\mu$.
The minimum feasible grade is $\mu+1$, so the same field with its tight threshold $m_\F(\beta)=v(\mu+1)$ attains $U=B\exp(K(v)-v)<B$ by~\eqref{eq:single-exact} and~\eqref{eq:single-finite-gap}.

\subsection{Counting and exact sampling on the same chain}
The supplied exchange rank is two, so Theorem~\ref{thm:count-main} supplies a deterministic mass FPTAS. Exact sampling has a simpler implementation here. Under the uniform half slice, exactly one of positions $1,n$ is selected with probability $n/[2(n-1)]$. On this event the two endpoint orientations have grades differing by $n-1>2$, so at least half the event lies outside $|Y|\le1$. Complementation splits the entire outside mass evenly between its positive and negative sides. Thus
\[
 \Pr(Y\ge2)\ge\frac{n}{8(n-1)}>\frac18.
\]
Every such word is feasible. Draw a uniformly random original quota word by exact integer unranking and reject unless both rows pass. The accepted law is exactly uniform on the actual feasible family, with fewer than eight expected proposals and polynomial expected bit cost.

%% file: proofs/path.tex
\section{Amplification on an incidence path}\label{app:path}
Fix $0<\alpha<1$. We prove a deterministic one-query reduction for the second assertion of Theorem~\ref{thm:path-main}. Use the stencilled private chain from Appendix~\ref{app:single-chain}, with $n$ original labels, full source size $B=\binom n{n/2}$, private capacity $U_b$, and private feasible count $Z_b$. Its data have the following properties:
\begin{itemize}
\item On YES, $U_b=B$; on NO, $\log(B/U_b)>16/n^3$.
\item A private field with tight log regret at most $1/(100n^3)$ is decoded by the rational centered-norm test.
\item Each private row uniquely maximizes at the same top quota word $T$, and its top slack is strictly positive.
\item $Z_b>B/8$, and the private family has a deterministic counting FPTAS and an exact uniform sampler.
\item The family has two accepted and two rejected words with equal paired incidence sums, on both decision outcomes.
\end{itemize}
Strict positive top slack follows also from the original row's uniform mean being its threshold and from strict variation on the source.

\subsection{Construction and size}
Choose a fixed integer $b$ such that
\begin{equation}\label{eq:path-exponent}
 b-3>\alpha(b+4),
\end{equation}
and use $t=n^b$ identical private copies. Between copies $j$ and $j+1$ place a quota-one gate with one low label and $m=n^4$ high labels. All gate activities are one. The first row of a private block uses the preceding gate, if present; the second row uses the following gate, if present. The two endpoint rows remain unchanged. For a private inequality $a(X)\ge\theta$, let $a_{\max}=a(T)$ and replace a gated occurrence by
\begin{equation}\label{eq:path-gate}
 a(X)+(a_{\max}-\theta)\ind_{\mathrm{high}}\ge a_{\max}.
\end{equation}
All rows remain commonly ordered: private scores increase strictly, and gate scores are zero on the low and the same positive number on each high. A high gate restores its incident private inequality. A low gate forces the adjacent private word to be $T$, by uniqueness of the maximum. Thus every feasible private tuple lies in $\F_b^t$, and every tuple in $\F_b^t$ is feasible when all gates are high. A low gate is allowed precisely when both neighboring private words are top.

The graph order is
\[
 a_1,\ C_1,\ b_1,\ E_1,\ a_2,\ C_2,\ b_2,\ E_2,\ldots,
 E_{t-1},\ a_t,\ C_t,\ b_t,
\]
with an edge exactly between consecutive vertices. Hence the row--chain graph is one path. Each private chain has local rank two and each gate has local rank one. The total original-label count is
\begin{equation}\label{eq:path-size}
 N=tn+(t-1)(m+1)=\Theta(tn^4).
\end{equation}
For fixed $\alpha$ the exponent $b$ is fixed. The labels, rows, and binary coefficients are all writable in time polynomial in the partition input length.

\subsection{A sandwich for capacity and private regrets}
Write $\beta_j$ for a joint field's restriction to private block $j$ and $\gamma_e$ for its gate coordinates. The gate coordinates may differ across high labels and have arbitrary signs. Restrict the full-source partition sum and the feasible minimum to the all-high face. The partition restriction decreases the numerator, while the support minimum on the face is at least the full support minimum. The face feasible family is the private product and the complete high domains. Therefore the joint tight log cost satisfies
\begin{align}
 G_{\rm path}(\beta,\gamma)
 &\ge \sum_{j=1}^tG_b(\beta_j)+
 \sum_{e=1}^{t-1}
 \left(\log\sum_{r=1}^m e^{\gamma_{er}}-\min_{1\le r\le m}\gamma_{er}\right)\notag\\
 &\ge \sum_{j=1}^tG_b(\beta_j)+(t-1)\log m.
 \label{eq:path-field-lower}
\end{align}
This allows arbitrary signed fields that distinguish the high labels. Conversely choose zero gate fields and private near-optimal fields. Every feasible private tuple lies in the private product, and the all-high face attains the sum of its separate support minima. The full partition still factors over all original gate labels. Taking infima yields
\begin{equation}\label{eq:path-capacity}
 m^{t-1}U_b^t\le U_{\rm path}\le(m+1)^{t-1}U_b^t.
\end{equation}
The argument does not require finite attainment of a private optimum.

Suppose a bare joint field is $\rho$-approximate. Combining~\eqref{eq:path-field-lower} and~\eqref{eq:path-capacity} gives
\begin{equation}\label{eq:path-regrets}
 0\le\sum_{j=1}^t\bigl(G_b(\beta_j)-\log U_b\bigr)
 \le\log\rho+(t-1)\log(1+1/m).
\end{equation}
Every summand is nonnegative.

\subsection{Decoding a single oracle answer}
Set $\rho=2^{N^\alpha}$. Equations~\eqref{eq:path-exponent} and~\eqref{eq:path-size} imply
\[
 \log\rho=o(t/n^3),\qquad
 (t-1)\log(1+1/m)\le t/n^4=o(t/n^3).
\]
For all sufficiently large $n$,~\eqref{eq:path-regrets} leaves fewer than $t/2$ private blocks with regret greater than $1/(100n^3)$. Apply the exact rational norm comparison from Appendix~\ref{app:single-chain} to every private restriction and take the majority. At least a strict majority give the correct partition answer, even though the oracle can choose different fields on different copies. A certified upper output supplies the same bare field.

For a symmetric value output use the integer reference $R=m^{t-1}B^t$ and the rational cutoff
\[
 R\,2^{-\lfloor t/n^3\rfloor}.
\]
On YES, $V\ge R/\rho$. On NO,
\[
 V\le R\rho\exp\bigl((t-1)\log(1+1/m)-16t/n^3\bigr).
\]
For large $n$ these lie strictly on opposite sides of the cutoff: its logarithmic offset is $(\log2)t/n^3+O(1)$, which dominates $\log\rho$ and is smaller than the NO loss after gate overhead. The cutoff numerator and denominator have polynomial bit length. The decoder need not approximate the real factor $2^{N^\alpha}$. The finitely many smaller item counts can be decided by enumeration, which is polynomial in the binary input length at a fixed item count. One query therefore gives the claimed NP-hardness for each contract.

\subsection{Minimum rank of the full path}
\begin{lemma}[Product trades force additive rank]\label{lem:product-trades}
Suppose each of $h$ nonempty block families has two rejected words and two accepted words whose paired incidence sums agree. Every exact common-order representation of their Cartesian product on the same quota source has exchange rank at least $2h$.
\end{lemma}
\begin{proof}
Consider any such representation and use its own common orders. Let $T_j$ be the simultaneous top word in block $j$. The top tuple is feasible by monotonicity. For any replacement row let $d_{js}\ge0$ be its drop from $T_j$ to rejected word $s=1,2$. The trade implies that the largest drop over accepted words in block $j$ is at least $(d_{j1}+d_{j2})/2$. The product allows these largest feasible drops to be attained simultaneously.

For each $(i,r)$ choose a row rejecting the tuple having rejected word $r$ only at block $i$, with all other blocks top. Its allowed drop budget is at least the sum of the blockwise maximum feasible drops, yet is strictly smaller than $d_{ir}$. Hence
\[
 d_{ir}>\tfrac12\sum_{j,s}d_{js},
 \quad\text{equivalently}\quad
 d_{ir}>\sum_{(j,s)\ne(i,r)}d_{js}.
\]
The resulting \(2h\)-square drop matrix is strictly diagonally dominant
and hence nonsingular. It factors as \(A'D\), where \(A'\) contains the
chosen rejecting rows and each column of \(D\) is the indicator of a top
word minus that of its rejected word, extended by zero outside its block.
Each such difference sums within-chain exchanges. Their images under
\(A'\) therefore span at most the exchange rank of the replacement
representation, for any replacement orders.
\end{proof}

Fix one particular high label at every gate. The restricted feasible family is the private product. Restricting any equivalent representation subtracts constants and removes fixed-chain exchange columns, so cannot increase rank. The private trades and Lemma~\ref{lem:product-trades} give the lower bound $2t$. The displayed $2t$ rows give the reverse bound. Hence the path's minimum representation rank is exactly $2t$ on both outcomes.

\subsection{A deterministic FPTAS for the true path count}
Let a private state $s_j=1$ indicate the unique top word, and $s_j=0$ a non-top feasible word. Their multiplicities are $w(1)=1$ and $w(0)=Z_b-1$. Between two states there are $m+s_js_{j+1}$ permitted original gate labels. Consequently
\begin{equation}\label{eq:path-count}
 Z_{\rm path}=
 \sum_{s\in\{0,1\}^t}\prod_{j=1}^tw(s_j)
 \prod_{j=1}^{t-1}(m+s_js_{j+1}).
\end{equation}
Let $f_j(s)$ count feasible labelled prefixes ending in private state
$s$, including the first $j$ private blocks and their $j-1$ internal
gates. The transfer is
\[
 f_1(s)=w(s),\qquad
 f_{j+1}(s')=w(s')\sum_{s\in\{0,1\}}(m+ss')f_j(s),
 \qquad Z_{\rm path}=f_t(0)+f_t(1).
\]
The only new choices are the next private word and the joining gate,
which proves this invariant by induction. Thus $O(t)$ arithmetic
steps evaluate the positive polynomial once $Z_b$ is known. Since $n\ge16$, the density bound gives $Z_b>B/8\ge2$. Use the private fixed-rank FPTAS to obtain $Z_b/(1+\eta)\le\widehat Z_b\le Z_b$, where $\eta=\varepsilon/(20t)$ and $0<\varepsilon\le1$. Then
\[
 (1-2\eta)(Z_b-1)\le\widehat Z_b-1\le Z_b-1.
\]
After substituting $\widehat Z_b$ for $Z_b$, each term retains at least a $(1-2\eta)^t$ fraction of its original value, where $(1-2\eta)^t\ge1-\varepsilon/10\ge1/(1+\varepsilon)$. The resulting transfer output is the desired one-sided approximation. Its denominators are powers of the one rational input estimate of degree at most $t$; all intermediate bit lengths are polynomial. This algorithm uses a rank-two private oracle, not a fixed-global-rank oracle on the path.

\subsection{Exact recursive sampling}
For an interval $I$, let $\F_I$ be its family on the original private and internal gate labels: each private word lies in $\F_b$, and a low internal gate requires both neighboring words to be $T$. An absent external gate is treated as high, so its exposed endpoint row uses the original private threshold. Thus both private feasibility tests remain on every block.

Use fresh independent fair bits in every call and restarted trial:
\begin{enumerate}
\item If $I$ consists of one block, return a word from the exact uniform private rejection sampler of Appendix~\ref{app:single-chain}.
\item Otherwise split $I$ at a middle gate into $I_L,I_R$. Independently generate one feasible labelled word on each child and draw a uniform original label of the joining gate.
\item Accept if the gate is high, or if both adjacent private words are top. Otherwise discard both child outputs and the gate label, and restart step 2 at this node.
\end{enumerate}
Inductively the two child outputs are independent and uniform on $\F_{I_L}$ and $\F_{I_R}$, even conditional on the completed history before the trial. Each candidate tuple therefore has probability
\[
 \frac{1}{|\F_{I_L}|\,|\F_{I_R}|\,(m+1)}.
\]
The accepted tuples are in bijection with $\F_I$: all private row tests are already enforced, and the only missing condition is exactly the joining gate test. Conditioning on acceptance gives the uniform law on the \emph{original labelled} parent words. The probability $a_I$ of acceptance is at least $m/(m+1)\ge1/2$, since every high label succeeds. This also proves almost-sure termination, by induction on interval length.

Write \(T(I)\) for expected bit cost and \(h(I)\) for the mean joining,
testing and copying cost per trial. A fresh trial has finite mean cost
\(c_I=T(I_L)+T(I_R)+h(I)\) by induction. Trial \(\ell\) is reached
with probability \((1-a_I)^{\ell-1}\), independently of its own cost.
Summing the nonnegative costs gives
\(T(I)=\sum_{\ell\ge1}(1-a_I)^{\ell-1}c_I=c_I/a_I\le2c_I\).
A balanced recursion has depth at most $\lceil\log_2t\rceil$, and the sum of these bounds has $O(t^2)$ node contributions. Leaf rejection costs fewer than eight expected trials. Exact uniform subset and gate-label generation, row tests, and copying each have polynomial bit cost; subset unranking uses the integer $\binom n{n/2}$, and fair-bit rejection generates every required uniform integer exactly. Thus the total expected bit time is polynomial. No private count, approximate transition probability, or independence of the two children \emph{after} the joining acceptance is required.

%% file: proofs/disclosure.tex
\section{Research use of generative AI}\label{app:ai-use}
ChatGPT and Codex were used extensively during the development of this research, including mathematical exploration, proposed constructions and counterexamples, proof derivation and revision, internal proof checking, literature search and comparison, and English drafting. They also assisted with exact finite computational checks, source integration, LaTeX preparation, and document verification.

This manuscript was assembled from a frozen archive of the existing mathematical results. Its preparation included separate AI-assisted reconstruction of the counting and capacity algorithms, output-sensitive hardness arguments, forest recurrences, shell certificates, and path amplification. The declared prior-work comparisons were checked against primary sources. Finite checks served as diagnostic evidence and did not replace the included general proofs.

These internal checks are not proof-assistant formalizations or independent human peer review. They do not certify priority or publication suitability. The human authors remain responsible for the submitted claims, source attribution, the accuracy of the disclosure, and the final submission. This statement does not assert that every proof has already been checked by a human author.

%% file: references.bib
@inproceedings{StraszakVishnoi2017,
 author={Damian Straszak and Nisheeth K. Vishnoi},
 title={Real Stable Polynomials and Matroids: Optimization and Counting},
 booktitle={Proceedings of the 49th Annual ACM SIGACT Symposium on Theory of Computing},
 year={2017}, pages={370--383}, doi={10.1145/3055399.3055457},
 eprint={1611.04548}, archivePrefix={arXiv}, url={https://arxiv.org/abs/1611.04548v1},
 note={Full version: \url{https://arxiv.org/abs/1611.04548v1}}
}

@inproceedings{StraszakVishnoi2019,
 author={Damian Straszak and Nisheeth K. Vishnoi},
 title={Maximum Entropy Distributions: Bit Complexity and Stability},
 booktitle={Proceedings of the Thirty-Second Conference on Learning Theory},
 series={Proceedings of Machine Learning Research}, volume={99},
 pages={2861--2891}, year={2019}, url={https://proceedings.mlr.press/v99/straszak19a.html}
}

@inproceedings{GopalanEtAl2011,
 author={Parikshit Gopalan and Adam R. Klivans and Raghu Meka and Daniel {\v S}tefankovi{\v c} and Santosh S. Vempala and Eric Vigoda},
 title={An {FPTAS} for {\#Knapsack} and Related Counting Problems},
 booktitle={Proceedings of the 52nd Annual IEEE Symposium on Foundations of Computer Science},
 pages={817--826}, year={2011}, doi={10.1109/FOCS.2011.32},
 url={https://www.microsoft.com/en-us/research/wp-content/uploads/2016/02/Knapsack-FOCS-final.pdf}
}

@inproceedings{Dyer2003,
 author={Martin Dyer}, title={Approximate Counting by Dynamic Programming},
 booktitle={Proceedings of the 35th Annual ACM Symposium on Theory of Computing},
 pages={693--699}, year={2003}, doi={10.1145/780542.780643},
 url={https://www.math.cmu.edu/~af1p/Teaching/MCC17/Papers/knapsack.pdf}
}

@inproceedings{GopalanKlivansMeka2012,
 author={Parikshit Gopalan and Adam R. Klivans and Raghu Meka},
 title={Learning Functions of Halfspaces Using Prefix Covers},
 booktitle={Proceedings of the 25th Annual Conference on Learning Theory},
 series={Proceedings of Machine Learning Research}, volume={23},
 pages={15.1--15.10}, year={2012}, url={https://proceedings.mlr.press/v23/gopalan12.html}
}

@inproceedings{AwerbuchKleinberg2004,
 author={Baruch Awerbuch and Robert D. Kleinberg},
 title={Adaptive Routing with End-to-End Feedback: Distributed Learning and Geometric Approaches},
 booktitle={Proceedings of the 36th Annual ACM Symposium on Theory of Computing},
 pages={45--53}, year={2004}, doi={10.1145/1007352.1007367},
 url={https://www.cs.cornell.edu/~rdk/papers/OLSP.pdf}
}

@inproceedings{PatelKhanLouis2021,
 author={Deval Patel and Arindam Khan and Anand Louis},
 title={Group Fairness for Knapsack Problems},
 booktitle={Proceedings of the 20th International Conference on Autonomous Agents and Multiagent Systems},
 pages={1001--1009}, year={2021},
 publisher={International Foundation for Autonomous Agents and Multiagent Systems},
 url={https://archive.illc.uva.nl/AAMAS-2021/pdfs/p1001.pdf},
 eprint={2006.07832}, archivePrefix={arXiv}
}

@inproceedings{Halman2016Forest,
 author={Nir Halman},
 title={A Deterministic Fully Polynomial Time Approximation Scheme for Counting Integer Knapsack Solutions Made Easy},
 booktitle={Approximation, Randomization, and Combinatorial Optimization. Algorithms and Techniques},
 series={Leibniz International Proceedings in Informatics}, volume={60},
 pages={9:1--9:11}, year={2016}, doi={10.4230/LIPIcs.APPROX-RANDOM.2016.9},
 url={https://drops.dagstuhl.de/entities/document/10.4230/LIPIcs.APPROX-RANDOM.2016.9}
}

@article{KschischangFreyLoeliger2001,
 author={Frank R. Kschischang and Brendan J. Frey and Hans-Andrea Loeliger},
 title={Factor Graphs and the Sum-Product Algorithm},
 journal={IEEE Transactions on Information Theory}, volume={47}, number={2},
 pages={498--519}, year={2001}, doi={10.1109/18.910572}
}

@article{Sion1958,
 author={Maurice Sion}, title={On General Minimax Theorems},
 journal={Pacific Journal of Mathematics}, volume={8}, number={1},
 pages={171--176}, year={1958}, doi={10.2140/pjm.1958.8.171},
 url={https://msp.org/pjm/1958/8-1/pjm-v8-n1-p14-s.pdf}
}

@incollection{Karp72,
  author = {Richard M. Karp},
  title = {Reducibility Among Combinatorial Problems},
  booktitle = {Complexity of Computer Computations},
  editor = {Raymond E. Miller and James W. Thatcher},
  pages = {85--103},
  year = {1972},
  publisher = {Plenum Press},
  doi = {10.1007/978-1-4684-2001-2_9}
}

@inproceedings{SinghVishnoi2014,
 author={Mohit Singh and Nisheeth K. Vishnoi},
 title={Entropy, Optimization and Counting},
 booktitle={Proceedings of the 46th Annual ACM Symposium on Theory of Computing},
 pages={50--59}, year={2014}, doi={10.1145/2591796.2591803},
 url={https://arxiv.org/abs/1304.8108v1},
 note={Full version: \url{https://arxiv.org/abs/1304.8108v1}. Theorem and appendix numbers refer to this version}
}

@inproceedings{FengJin2025,
  author = {Weiming Feng and Ce Jin},
  title = {Approximately Counting Knapsack Solutions in Subquadratic Time},
  booktitle = {Proceedings of the 2025 Annual ACM-SIAM Symposium on Discrete Algorithms},
  year = {2025},
  doi = {10.1137/1.9781611978322.32}
}

@misc{JinParkSaha2026,
  author = {Ce Jin and Jaewoo Park and Barna Saha},
  title = {Faster Deterministic Approximation for Counting Knapsack},
  howpublished = {arXiv:2610.14595},
  year = {2026}
}
